\documentclass[journal]{IEEEtran}

\usepackage{amsmath,amsfonts}
\usepackage{algorithmic}
\usepackage{array}
\usepackage{textcomp}
\usepackage{stfloats}
\usepackage{url}
\usepackage{verbatim}
\usepackage{graphicx}
\usepackage{amsmath,amssymb,amsthm}
\usepackage{booktabs}
\usepackage{float}
\usepackage{dsfont}
\usepackage{xcolor}
\usepackage{enumitem}
\usepackage{soul}
\usepackage{arydshln} % 放在导言区
\usepackage{diagbox}
\usepackage{wasysym}
\usepackage{booktabs}
\usepackage{multirow}  % 你已经用了 multirow
\usepackage{makecell}

\usepackage{subcaption}
\usepackage{bm}

\usepackage[ruled,linesnumbered,vlined]{algorithm2e}

\definecolor{gblue}{rgb}{0, 0.3, 0.59}
\definecolor{purple}{rgb}{0.8, 0.2, 1}
\definecolor{blue}{rgb}{0, 0, 1}

\newtheorem{remark}{Remark}

\newtheorem{example}{Example}

\allowdisplaybreaks[4]

\usepackage[colorlinks, linkcolor=gblue, anchorcolor=gblue, citecolor=gblue, urlcolor=gblue]{hyperref}

\usepackage[numbers,sort&compress]{natbib}

\begin{document}

\title{
% Anomaly Detection in Protocol Configurations: A Context-aware Approach for Wireless Network Resilience
% 
% Protocol configuration anomaly Detection for Network Resilience: A Configuration and Neighbor Context-Aware GAT Approach
% Agentic AI for Network Resilience: Detecting Anomalous Configurations in Sovereign Network Functions
% GSID: 
Enhancing Network Resilience via Graph-Based Anomaly Detection in Sovereign Functions
% Configuration- and Neighbor-Aware GNN: 
% 
}
% highlight the dynamic and adaptiveness in this paper.

\author{
Xin~Hao, 
Wei~Ni,~\IEEEmembership{Fellow,~IEEE},
Chenhan~Zhang,
Massimo~Piccardi,
and~Raymond~Owen
% % 
\thanks{
This work was supported by the {Department of Infrastructure, Transport, Regional Development and Communications under Grant SON3352211}. 
This article was accepted in part at the 2026 IEEE Global Communications Conference (GLOBECOM)~\cite{Xin_GC_2026}.
\emph{(Corresponding author: W. Ni.)}
}
% % 
\thanks{
X. Hao, W. Ni, C. Zhang, M. Piccardi, and R. Owen are with the Faculty of Engineering and Information Technology, University of Technology Sydney, NSW, Australia (e-mail: \{xin.hao; wei.ni; chenhan.zhang; massimo.piccardi; ray.owen\}@uts.edu.au).
}}
\markboth{IEEE Transactions on Network Science and Engineering} %
{Shell \MakeLowercase{\textit{et al.}}: Bare Demo of IEEEtran.cls for IEEE Journals}

\maketitle

% graph structural inconsistency detector (GSID) for communication networks

\begin{abstract}
Sovereign network functions, e.g., routing protocols, are becoming increasingly complex and susceptible to failures arising from protocol configuration anomalies and anomalous configurations.
This paper interprets the protocol configuration anomaly detection problem as detection of structural inconsistencies of connected nodes and edges in a bipartite graph that captures both physical network entities and logical protocol states. This graph structural inconsistency detector (GSID) model is proposed to solve the problem efficiently. To handle the heterogeneous nature of protocol configuration parameters, GSID employs an adaptive configuration encoder (ACE) that dynamically selects encoding strategies per parameter to preserve fine-grained numerical discrepancies. To expose the subtle inconsistencies of connected nodes and edges in the bipartite graph, GSID uses an inconsistency dynamic attention (IDA) mechanism that scores edges by drawing asymmetric attentions from both ends, rule compliance from one end, and route connectivity from the other. 
It is demonstrated experimentally that GSID outperforms state-of-the-art baselines by threefold in F1 score and by $23.2\%$ in accuracy. Ablation studies validate the effectiveness of both the ACE and IDA modules. {Out-of-distribution (OOD) generalization tests} on unseen network scales and real-world network topologies show the superior adaptability of GSID, compared to the baselines.
\end{abstract}
% inconsistency dynamic attention

\IEEEpeerreviewmaketitle

% \subsection{Motivations}

\section{Introduction}\label{section_introduction}
\IEEEPARstart{W}{ith} the fast expansion and evolution of communication networks and protocols, e.g., 4G/5G, the networks are increasingly complex and prone to risks~\cite{Survey_contemporary_resilience_2025}, such as protocol configuration anomalies and hardware failures~\cite{oecd2025_misconf, Magazine_Turkey_earthquake, tii_link_fault_IoT}. In particular, 4G and 5G (and equally likely future 6G) network operations rely on a large number of sovereign functions, 
% {which are the Internet Protocol (IP) functions that orchestrate internet traffic flows coordinated internally, at the edge, and externally~\cite{Ray_Sovereign_Functions}. Typical sovereign functions include} Border Gateway Protocol (BGP), Open Shortest Path First (OSPF), and Multiprotocol Label Switching (MPLS). These functions and protocols 
which orchestrate routing decisions, manage traffic engineering, and maintain connectivity across autonomous domains, serving as the backbone that underpins the reliability of modern communication networks~\cite{Ray_Sovereign_Functions}. 
% \gray{However, they can be prone to configuration anomalies due to dramatically increased complexity resulting from the expansion and evolution of networks~\cite{oecd2025_misconf}, and the configuration anomalies can be caused by human errors and cyber attacks~\cite{BGP_anomaly_survey, GraphBGP, Xin_DBC}.}
%%% Please incorporate the above references into the new paragraph (last few sentences)

Sovereign functions are the critical IP functions whose configuration governs the routing, management, and security of traffic across a network, coordinated internally, at the edge, and externally, and on which nation-scale service continuity depends. They span routing and path determination, e.g., Border Gateway Protocol (BGP), Open Shortest Path First (OSPF), and Multiprotocol Label Switching (MPLS); addressing and naming, e.g., Domain Name System (DNS) and Dynamic Host Configuration Protocol (DHCP); time synchronization, e.g., Network Time Protocol (NTP) and Precise Time Protocol (PTP); and the traffic-filtering and access-control policies, e.g., firewall rules and Access Control Lists (ACLs) that enforce them. In each, a configuration can be locally valid yet semantically inconsistent, so a single misconfiguration may cascade network-wide without raising an explicit fault~\cite{BGP_anomaly_survey, GraphBGP, Xin_DBC}. This paper addresses this class of protocol configuration anomalies, focusing on routing, specifically BGP and OSPF, as the representative and highest-impact case.

Several notable examples of recent network failures or malfunctions include the 2024 AT\&T misconfigured network element~\cite {2024_ATT_outage}, 2022 KDDI misconfigured router route~\cite{2022_KDDI_outage}, and 2020 T-Mobile configuration anomalies~\cite{2020_T_mobile_outage}, where the post-failure diagnosis indicated the sovereign functions were not adequately configured during network upgrade~\cite{Ray_Sovereign_Functions}. Even when each parameter of individual network entities is configured inadequately but still within the tolerable range of the parameter configurations~\cite{SIGCOMM_Understand_BGP_Misconfig}, the cascade of this inadequacy eventually leads to an impact on the entire network~\cite{nature_cascading_failure_effects, TNSE_cascading_tutorial}.

Artificial intelligence (AI) is increasingly demonstrating its capability and potential in network resilience, e.g., for adaptive and scalable frameworks for 6G communications~\cite{Feibo_JSAC_agenticAI, liuchang_2022learning}, fault-tolerant service migration in edge computing~\cite{TNSE_agenticAI}, and enabling end-to-end multi-task orchestration via highly scalable graph neural networks (GNNs)~\cite{WEI_COMST_Agentic_GNN, AgenticGNN_JSAC_dusit}. To date, AI has not been considered for detection and prevention of network protocol configuration anomalies in sovereign functions, although it can be uniquely suited due to its collaborative perception across distributed network entities, reasoning over heterogeneous protocol states, and adaptivity to dynamic and previously unseen network scenarios. {Moreover, a well-designed graph-based representation can inherently unify multiple sovereign functions within a single framework, enabling joint anomaly detection across heterogeneous protocols without protocol-specific retraining.}
% , AI agents can be deployed across the networks, collect the update of sovereign function protocol parameters, create situational awareness, and alert emerging risks or inadequate configurations that may cascade over networks. 

\subsection{Challenges}
Despite its criticality, it is non-trivial to detect configuration anomalies or inadequate configurations of sovereign functions due to the subtlety and sophistication of the configurations~\cite{ComST_BGP_anomaly_detection, Xin_ISWCS}. As mentioned earlier, an inadequate configuration of the sovereign functions at a network entity is not necessarily incorrect, but can potentially cascade across the network. For example, incorrect configuration of destinations may get traffic to go nowhere without triggering any explicit alarm~\cite{SIGCOMM_Understand_BGP_Misconfig}. On the other hand, network entities, e.g., routers and gateways, may have to comply with rules like mutual accessibility or exclusion, for security purposes~\cite{acma2025optus}. While verifying the legitimacy of a protocol configuration, the collective compliance of all network entities involved also needs to be assessed at the cost of combinatorial complexity. 

Another crucial challenge of the detection and prevention of sovereign function configuration anomaly or their inadequate configuration is that spatial variation (and locality) often exists when such configuration anomaly or inadequate configuration occurs~\cite{CloudTrie}. In particular, a configuration of a sovereign function, e.g., BGP, can impact any network entities associated with the function or protocol, e.g., along the route specified by the configuration~\cite{ComMag_BGP_cascade}. Moreover, the impact can potentially propagate beyond, because some of the network entities can also be engaged in other sovereign functions.

While GNNs are suited to intelligent decision-making for network management~\cite{WEI_COMST_Agentic_GNN, TNSE_agenticAI, Feibo_JSAC_agenticAI}, the specialized designs may still be challenging when considering the stringent requirement of low complexity, high scalability, and quick adaptivity for fast detection of inadequate configurations of sovereign functions and protocols. It is critical to produce fast detection of inadequate configurations for the sake of early prevention of significant consequences, e.g., those seen in~\cite{2024_ATT_outage, 2022_KDDI_outage, 2020_T_mobile_outage}.

\subsection{Contributions}
This paper presents a new graph structural inconsistency detector (GSID) for protocol anomalies (i.e., configuration anomalies and/or inadequate configurations of protocols in sovereign functions) in existing and emerging, large-scale communication networks, with the following contributions:
\begin{enumerate}
\item 
We interpret the protocol configuration anomaly detection problem graph-theoretically after constructing a bipartite graph that encodes the relationships among protocol states, network entities, and rule compliance. Anomalous configurations are identified as structural and semantic inconsistencies within edges capturing entity involvement and compliance constraints, and vertices representing protocol configurations.
\item We propose a GNN-based algorithm, termed GSID, to effectively detect inconsistencies in the bipartite graph. This is achieved through a tailored encoding strategy and a dynamic attention mechanism that captures structural inconsistencies in the graph representations.

\item We design an ACE that adaptively selects the appropriate encoding strategy according to the characteristics of configuration parameters, providing subsequent GNNs with fine-grained numerical discrepancies for detecting protocol anomalies. 
\item We propose an IDA mechanism, which detects protocol anomalies by capturing the asymmetric influence from both ends of edges in the bipartite graph. Attention is drawn to complementarily account for rule compliance on routes from one end and physical connectivity along routes from the other end, hence helping reveal subtle inconsistencies stemming from anomalies.

\item {The bipartite graph representation is protocol-agnostic by design: new sovereign functions beyond BGP and OSPF can be incorporated by defining the corresponding predicate types and edge structures without modifying the ACE or IDA modules, enabling GSID to operate across multi-protocol scenarios under a unified representation.}
\end{enumerate}

Extensive experiments are conducted to validate GSID. Ablation studies confirm the individual contributions of the ACE and IDA modules. It is demonstrated that GSID outperforms the state-of-the-art baseline by more than 3 times in F1 score and $23.2\%$ in accuracy. {Out-of-distribution (OOD) generalization testing on unseen network scales, real-world network topologies from the Internet Topology Zoo~\cite{TopologyZoo}, and unseen anomaly injection rates validate the superior adaptability of GSID, compared to the baselines, where the protocol configuration parameters are synthetically generated with the topology structures derived from operational networks.}

The rest of this paper is organized as follows. Section \ref{section_related_works} summarizes the related works. Section~\ref{section_system_model} presents the system model and formulates the protocol configuration anomaly detection problem. Section~\ref{section_proposed_algorithm} elaborates the proposed GSID, including the CA node feature encoder, the IDA mechanism, the training procedure, and the computational complexity analysis. Section~\ref{section_performance_evaluation} reports the training and testing results. Section~\ref{section_conclusion} summarizes the paper and future directions.

\section{Related Works}\label{section_related_works}
Table~\ref{tab_related_works} summarizes representative studies related to protocol configuration anomaly detection, highlighting their research categories, network representations, technical approaches, and challenges addressed.

\subsection{Network Resilience and Protocol Configuration}\label{subsec_resilience}

Communication network outages have occurred worldwide with alarming frequency~\cite{Magazine_Turkey_earthquake, ComMag_resilience_eng}. These outages make network resilience, i.e., the ability to provide services despite network degradations and failures~\cite{definition_resilience_D2R2, inet_resilient_and_robust}, ranging from physical infrastructure failures to logical configuration errors~\cite{Survey_contemporary_resilience_2025, Ray_Sovereign_Functions}. Much research attention has been devoted to restoring network outages caused by physical failures, i.e., network links and entities~\cite{inet_single_link_failure, inet_link_failure_IoT_mesh, inet_server_failure}. As modern networks become increasingly complex and semantically rich~\cite{ComMag_AI_IBN, NetKG}, knowledge-driven and autonomous management mechanisms have been proposed to cope with evolving operational conditions~\cite{Google_MALT, knowledge_driven_autonomous_network}. However, none of these advances focus on detecting protocol configuration anomalies in sovereign functions, which have become the dominant cause of network outages in recent years~\cite{oecd2025_misconf}.

Sovereign network functions, such as BGP and OSPF, govern how traffic traverses autonomous systems (ASes) and intra-domain topologies, and their configurations are central to network operations~\cite{Ray_Sovereign_Functions}. These configurations have grown increasingly complex with the expansion of modern networks, involving a large number of parameters that jointly determine routing decisions across multiple administrative domains~\cite{rfc4271, rfc_OSPF}. BGP alone exposes dozens of configurable attributes, including local preference, AS path length, multi-exit discriminator (MED), and community values~\cite{rfc4271}, while OSPF introduces link weights that govern intra-AS shortest-path computation~\cite{rfc_OSPF}. Compounding this complexity, the topologies of contemporary 4G/5G networks~\cite{StandMag_routing_protocol, TNSE_SFC} evolve continuously due to channels being established and torn down~\cite{TNSE_TOAST, TNSE_dynamic_channel_multi_agent}, devices joining and leaving~\cite{TNSE_dynamic_device_authentication, Xin_BCDRL}, and routing relationships adapting to varying service demands~\cite{TNSE_routing_dynamic, Xin_ICC2024}. Configurations must be updated adaptively to remain aligned with the current service requirements; any inconsistency between a configuration and its operational context can manifest as an anomaly~\cite{APGNN, TII_alarm_flood_1, TASE_alarm_flood_2}, even when the configured values remain valid.

A notable characteristic across network protocols is the heterogeneity of different configuration parameters. Some parameters, like BGP local preference, are numerical quantities whose routing consequences depend on their relative magnitudes, where even small perturbations can cascade into network-wide route changes~\cite{ComMag_BGP_cascade, TNSM_GonoGo}. Some other parameters, like BGP origin, are categorical in nature and serve as discrete identities~\cite{cisco_base_configuration}, so any deviation represents a change of identity. The coexisting numerical and categorical parameters jointly determine routing outcomes~\cite{NeurIPS_BGP}, calling for encoding strategies that respect their distinct characteristics.

% \textit{Lessons Learned:}
% \begin{itemize}
%     \item Network outages in modern communication networks are largely caused by configuration anomalies in Sovereign network functions.
%     \item The heterogeneity of configuration parameters demands adaptive encoding strategies to expose anomaly signals that uniform encoding would overlook.
%     \item The growing complexity of modern networks calls for detection frameworks that can reason jointly over configurations, topology, and service requirements.
% \end{itemize}

{The lessons learned from the above-mentioned existing studies are that configuration anomalies in sovereign network functions are the dominant cause of modern network outages, and their heterogeneous parameter characteristics demand adaptive, joint encoding strategies that reason over configurations, topology, and service requirements simultaneously.}

\subsection{Graph Representation of Communication Networks}\label{subsec_graph_representation}
Communication networks are inherently graph-structured, with devices and their interconnections directly mapping to nodes and edges. This has motivated extensive use of graph representations in networking modeling. Conventional graph representations typically encode physical topology, where nodes correspond to network devices and edges correspond to physical links~\cite{RouteNet, RouteNet_fermi, RouteNet_Gauss, Xin_ICC2023}. Such representations have been used successfully for topology-centric tasks, e.g., routing optimization~\cite{TNSE_routing_optimization, TNSE_routing_satellite}, traffic engineering~\cite{ZCH_FASTGNN, TNSE_traffic_engineering}, and resource allocation~\cite{Xin_HML, Yuhong_TVT_GNN}, where the primary concern is how signals or flows traverse the physical infrastructure. However, none of these works encode protocol states, routing intents, or compliance rules in the graphs, which are essential for capturing the routing behaviors shaped by protocol configurations.

Bipartite and heterogeneous graph representations have been proposed to unify distinct types of network entities within a single graph, e.g., for wireless network use accessing~\cite{LXM_GNN_GC_2021}, and operational fix network modeling~\cite{Google_MALT} and management~\cite{Google_topoplan}. While the study~\cite{Google_MALT} incorporates operational semantics into the heterogeneous graph for management purposes, neither of these uses targets protocol configuration anomaly detection.
% \textit{Lessons Learned:}
% \begin{itemize}
%     \item Conventional topology-only graph representations overlook the encoding of protocol states, routing intents, or compliance rules, and are insufficient for protocol configuration anomaly detection.
%     \item Existing bipartite and heterogeneous graph representations unify distinct entity types for tasks such as access modeling and network management, and the configuration semantics needed for anomaly detection remain to be incorporated.
% \end{itemize}
{The lessons learned are that existing graph representations, whether topology-only or bipartite, {have not incorporated} the protocol states, routing intents, and compliance rules essential for configuration anomaly detection.}

\subsection{Anomaly Detection in Communication Networks}\label{subsec_anomaly_detection}
Despite the representational advances discussed above, anomaly detection in operational networks has long relied on observed telemetry, where devices continuously emit large volumes of alarms, logs, and state-change notifications~\cite{ComST_rule_based}. Operators routinely face alarm floods~\cite{TII_alarm_flood_1, TASE_alarm_flood_2}, in which genuine anomalies are buried among vast quantities of benign or redundant events~\cite{Xin_WFIoT}; both rule-based and learning-based methods have been developed to surface real faults from such telemetry streams~\cite{ComST_ML_for_wireless, APGNN}. Beyond alarm-driven detection, a broader body of work flags anomalies as deviations from learned statistical patterns, applied to intrusions and malicious flows in communication networks~\cite{TNSM_intrusion_detection, ACM_SVM_malicious_usr}, device-level monitoring in industrial control systems~\cite{TNSE_ICS_anomaly_detection}, and transaction-level detection in blockchain networks~\cite{TNSE_blockchain_anomaly_detection}. These approaches rely on a common assumption that anomalies produce explicit symptoms, i.e., observable deviations in measurable signals~\cite{BGP_anomaly_survey}. However, protocol configuration anomalies violate this assumption, as the symptoms of such anomalies are often implicit. An anomalous configuration value is syntactically valid and fully within valid ranges, triggering no direct signal for telemetry-based methods to act on~\cite{TNSM_protocol_failure}. Its consequences instead propagate through the joint state of multiple network entities, eventually causing routing behaviors inconsistent with operational intent. Detecting such anomalies requires examining the configurations themselves, but existing approaches applied to historical configuration data, e.g., support vector machines~\cite{ACM_SVM_malicious_usr}, decision trees~\cite{JSAC_decision_tree_as_benchmark}, and feed-forward neural networks~\cite{DeepBGP}, treat parameters uniformly or independently, overlooking the structural dependencies among network entities that routing protocols inherently induce.

% \textit{Lessons Learned:}

% \begin{itemize}
%     \item Existing anomaly detection methods in communication networks predominantly target explicit symptoms, i.e., observable deviations in telemetry data. However, the symptoms of protocol configuration anomalies are implicit, propagating through joint states rather than surfacing as local signals, which such methods cannot capture.
%     \item Methods that directly target configurations treat parameters uniformly or independently, overlooking the structural dependencies among network entities that routing protocols inherently induce.
% \end{itemize}

Some important lessons learned from the existing studies are that existing anomaly detection methods, whether telemetry-driven or configuration-targeted, {are not designed to} detect protocol configuration anomalies because such anomalies produce no observable local symptoms and instead propagate implicitly through the structural dependencies among network entities.

\subsection{GNN-Based Configuration Anomaly Detection}\label{subsec_gnn_approaches}
GNN-based approaches are inherently fit for protocol configuration anomaly detection in communication networks, as their graph-native design~\cite{MPNN} is well-suited to processing the structural dependencies that routing protocols induce. Within this line of research, existing works vary primarily in how configuration parameters are encoded into node features and how these features are aggregated across the graph.
% , and we review the two stages in turn below.

Regarding the encoding stage, GNNs have been applied to BGP-related tasks with different encoding strategies. For instance, in~\cite{NeurIPS_BGP}, the authors used lookup table-based encoding for all the BGP configuration parameters in their GNN for the configuration synthesis task. In~\cite{BGP_GNN_community}, the authors proposed a GNN-based approach for BGP community attribute classification, where node features are constructed by concatenating AS degree statistics and relationship type indicators. In~\cite{BGP_GNN_anomaly}, the authors proposed a GNN-based model for BGP anomaly detection, where each node is represented by the count of originated prefixes. All these works~\cite{NeurIPS_BGP, BGP_GNN_community, BGP_GNN_anomaly} encode configuration parameters uniformly without distinguishing their characteristics, leaving the parameter heterogeneity highlighted in Section~\ref{subsec_resilience} unaddressed despite BGP involving a variety of parameters with fundamentally different natures~\cite{GraphBGP}.

Regarding the aggregation stage, how node features are aggregated across the graph fundamentally determines whether diagnostic signals can be effectively captured, as anomaly symptoms propagate non-uniformly through the network, with certain paths and nodes becoming disproportionately critical for identifying anomalous parameters. Conventional GNNs aggregate messages uniformly across all neighbors~\cite{GCN_classical, MPNN}, and graph attention mechanisms~\cite{GAT_ICLR_2018, GTN_NeurIPS} have been introduced to weight the importance of neighboring nodes during message passing. Several recent works have applied attention-based GNNs to network anomaly detection tasks, including attack detection in industrial networks via state-transition modeling~\cite{TNSE_SF_ADL}, multivariate time-series anomaly detection via dual graph learning~\cite{TNSE_DGLAD}, and hierarchical anomaly detection in 6G networks~\cite{TCCN_node_level_attention_6G}. However, none of these existing works~\cite{TNSE_SF_ADL, TNSE_DGLAD, TCCN_node_level_attention_6G} account for the fact that the diagnostic relevance of a neighboring node depends on the joint state of both endpoints rather than each endpoint independently, a dependency that is critical for protocol anomaly detection, where even a small configuration anomaly may trigger a network-wide outcome.

% ====================================================================
\begin{table*}[!t]
\centering
\caption{Summary of Representative Studies Related to Protocol Configuration Anomaly Detection.}
\label{tab_related_works}
\renewcommand{\arraystretch}{1.16}
\setlength{\tabcolsep}{4pt}
\small
\begin{tabular}{m{2.2cm} m{2cm} m{4.5cm} m{4cm} m{4cm}}
\toprule
\toprule
\textbf{Category} & \textbf{Refs} & \textbf{Network Representation} & \textbf{Technical Approach} & \textbf{Challenges Addressed} \\
\midrule

\multirow{3}{=}{\textbf{Outage Evidence and Resilience}}
& \cite{2024_ATT_outage, 2022_KDDI_outage, 2020_T_mobile_outage, acma2025optus} & Incident descriptions of networks & Post-failure diagnosis & Config-induced outages \\
\cmidrule(l){2-5}
& \cite{Magazine_Turkey_earthquake, ComMag_resilience_eng, Survey_contemporary_resilience_2025, definition_resilience_D2R2, inet_resilient_and_robust} & Physical topology & Resilience frameworks & Physical failure recovery \\
\cmidrule(l){2-5}
& \cite{oecd2025_misconf, BGP_anomaly_survey, ComST_BGP_anomaly_detection} & Aggregated outage statistics & Survey and statistics & Misconfig as outage cause \\
\midrule[\heavyrulewidth]

\multirow{3}{=}{\textbf{Protocol and Config Basics}}
& \cite{rfc4271, rfc_OSPF, cisco_base_configuration} & Protocol parameter specifications & Specification documents & Parameter semantics/ranges \\
\cmidrule(l){2-5}
& \cite{Ray_Sovereign_Functions} & Inter and intra AS topology & Conceptual framing & Sovereign function scope \\
\cmidrule(l){2-5}
& \cite{ComMag_BGP_cascade, SIGCOMM_Understand_BGP_Misconfig, nature_cascading_failure_effects, TNSE_cascading_tutorial} & AS-level routing graphs & Cascading effect analysis & Parameter-level misconfig cascades \\
\midrule[\heavyrulewidth]

\multirow{3}{=}{\textbf{Explicit-Symptom Detection}}
& \cite{ComST_rule_based, TII_alarm_flood_1, TASE_alarm_flood_2, APGNN, Xin_WFIoT} & Alarm and log event streams & Rule- and learning-based triage & Alarm flood reduction, fault surfacing \\
\cmidrule(l){2-5}
& \cite{TNSM_intrusion_detection, ACM_SVM_malicious_usr} & Traffic flow statistics & Statistical deviation flagging & Intrusion, malicious flow detection \\
\cmidrule(l){2-5}
& \cite{TNSE_ICS_anomaly_detection, TNSE_blockchain_anomaly_detection} & Device states, transaction records & Behavioral baseline deviation & ICS, blockchain anomaly detection \\
\midrule[\heavyrulewidth]

\textbf{Config-Targeted Classical ML}
& \cite{ComST_ML_for_wireless, TNSM_GonoGo, NetKG, DeepBGP, JSAC_decision_tree_as_benchmark} & Per-node config feature vectors & SVM, decision tree, feed-forward NN & BGP, wireless config classification \\
\midrule[\heavyrulewidth]

\multirow{2}{=}{\textbf{Graph Representations}}
& \cite{RouteNet, RouteNet_fermi, RouteNet_Gauss, Xin_ICC2023, TNSE_routing_optimization, TNSE_routing_satellite, ZCH_FASTGNN, TNSE_traffic_engineering, Xin_HML, Yuhong_TVT_GNN} & Physical topology graphs (nodes: devices, edges: links) & GNN message passing & Routing, traffic engineering, resource allocation \\
\cmidrule(l){2-5}
& \cite{LXM_GNN_GC_2021, Google_MALT, Google_topoplan} & Bipartite, heterogeneous graphs & Type-aware aggregation & Wireless access, network management \\
\midrule[\heavyrulewidth]

\multirow{2}{=}{\textbf{GNN and Attention Foundations}}
& \cite{GCN_classical, MPNN} & Generic graphs & Uniform neighbor aggregation & Graph-native learning baseline \\
\cmidrule(l){2-5}
& \cite{GAT_ICLR_2018, GTN_NeurIPS, GATv2} & Generic attributed graphs & Attention-weighted aggregation & Non-uniform neighbor importance \\
\midrule[\heavyrulewidth]

\multirow{2}{=}{\textbf{GNN-Based Config and Anomaly Detection}}
& \cite{NeurIPS_BGP, BGP_GNN_community, BGP_GNN_anomaly, GraphBGP} & AS-level BGP graphs & Uniform parameter encoding with GNN & BGP synthesis, classification, anomaly \\
\cmidrule(l){2-5}
& \cite{TNSE_SF_ADL, TNSE_DGLAD, TCCN_node_level_attention_6G} & Domain-specific graphs (state-transition, sensor relation, 6G hierarchy) & Attention-based GNN aggregation & ICS attack, time-series anomaly, 6G anomaly \\

\midrule[\heavyrulewidth]
\textbf{Proposed GSID} 
& This work
& Bipartite graph (physical entities (routers, ASes, networks); semantics (protocol states, rules) 
& ACE (adaptive configuration encoder), IDA (inconsistency dynamic attention) 
& Protocol configuration anomaly detection \\

\bottomrule
\bottomrule
\end{tabular}
\end{table*}
% ====================================================================

The lessons learned from these existing studies are that existing GNN-based approaches uniformly encode heterogeneous configuration parameters and independently decompose endpoint contributions during aggregation, {and do not address} the fine-grained numerical discrepancies and joint-state dependencies that protocol anomaly detection requires.
{In particular, existing graph attention mechanisms, such as GATv2, are general-purpose and do not exploit the semantic roles of the two endpoint types in the bipartite graph, namely rule compliance from the fact node end and route connectivity from the entity node end, which are essential for exposing structural inconsistencies pertaining to protocol configuration anomalies.}

\begin{figure}[t]
    \centering
    \includegraphics[width=0.95\linewidth]{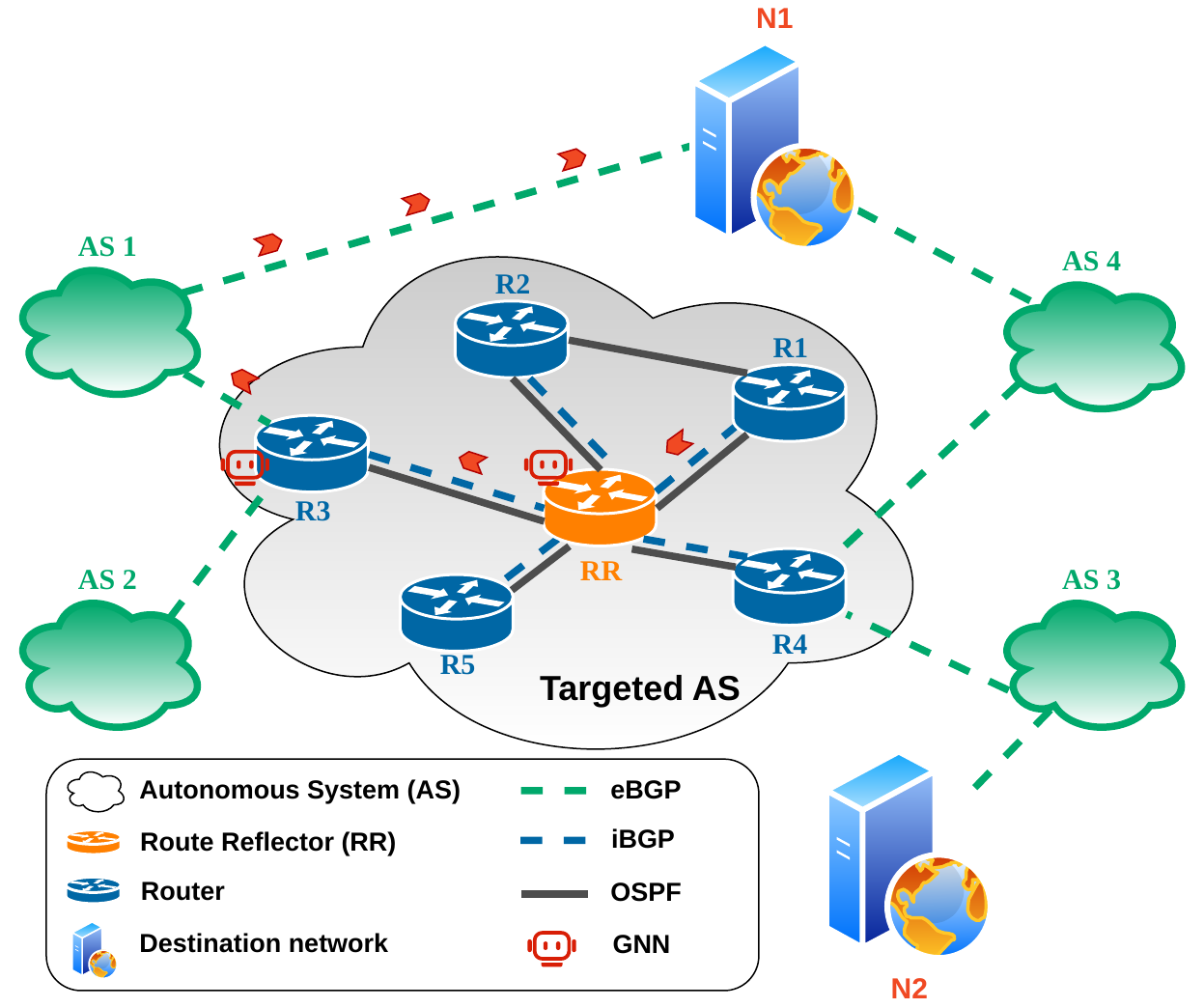}
    \caption{According to the network protocols, the data packets (red) from router R1 to destination network N1 are forwarded through router R3 along the OSPF-computed path (solid line), using the route learned by iBGP from R3, then advertised by eBGP from AS1.}
    % \caption{An example of routing in a BGP/OSPF network.}
    \label{fig_system_model}
\end{figure}
% =============================================================

\section{System Model and Problem Formulation}\label{section_system_model}
As depicted in Fig.~\ref{fig_system_model}, a communication network consists of multiple ASes with each comprising a set of routers interconnected by physical links. Within each AS, gateway routers (e.g., $R3$, $R4$) sit on the boundary of the AS and connect to neighboring ASes via direct physical links. Internal routers (e.g., $R1$, $R2$) are interconnected within the AS and communicate with a designated route reflector (RR). External destination networks (e.g., $N1$, $N2$) are reachable through the gateway routers.

\subsection{Network Protocol Configuration}\label{section_protocol_configuration_model}
As mentioned in Section~\ref{section_introduction}, the BGP and the OSPF are two of the most widely adopted routing protocols, governing traffic across the physical infrastructure of networks. 

% Path-Vector = 我不只告诉你能到达哪里，还告诉你完整的经过路径，让你既能防环，又能基于路径做灵活的策略决策。这正是 BGP 作为互联网核心路由协议的关键设计。
\subsubsection{BGP Configurations}
The BGP is a path-vector routing protocol.
Specifically, the external BGP (eBGP) runs between gateway routers of neighboring ASes, learning which external destination networks are reachable and through which AS.
The internal BGP (iBGP) runs among all routers within each AS through an RR, which redistributes these externally learned routes to all internal routers. 

The BGP selects the best route by comparing three configurable parameters comprising the BGP configuration set:
% denoted by
\begin{equation}
    \mathcal{C}^{\mathrm{BGP}} 
    \triangleq 
    \left\{\theta_{v}^{\mathrm{LP}}, ~ \theta_{v,u}^{\mathrm{ASL}}, ~ \theta_{v,u}^{\mathrm{MED}}, ~~ \forall v,u ~ (v \neq u) \right\},
    \label{eq_bgp_parameters}
\end{equation}
where $\theta_{v}^{\mathrm{LP}}$ is the local preference attached to the $v$-th router, specifying the preferred exit point within the AS, with higher values indicating higher priority; $\theta_{v,u}^{\mathrm{ASL}}$ is the AS path length of the route between routers $v$ and $u$, counting the number of ASes traversed, with shorter paths indicating higher priority; and $\theta_{v,u}^{\mathrm{MED}}$ is the MED advertised between routers $v$ and $u$, suggesting the preferred entry point, with lower values indicating higher priority.

The BGP also reflects networking intents that describe the rules and attributes of routes and/or routing requirements. For example, the BGP configurations control the actual routing paths among the routers, which are expected to be consistent with the preset networking intents, such as mutual reachability (i.e., whether specific nodes can reach each other) and traffic isolation (i.e., whether certain traffic flows remain separated). 
% it can specify the mutual reachability among a set of network nodes, and mutual exclusion or isolation among nodes. 
These descriptions of networking intents are preset. They provide references/rules for BGP routing, and stay unchanged during network operations.

\subsubsection{OSPF Protocol Configurations}
OSPF is a link-state routing protocol that runs on all routers within an AS, and is responsible for computing intra-AS forwarding paths. 
Each router independently executes Dijkstra's algorithm~\cite{Dijkstra} over the network topology to find the shortest path to every other router within the AS, with the path cost determined by the following OSPF configuration parameter
\begin{equation}
    \mathcal{C}^{\mathrm{OSPF}} \triangleq \left\{\phi_{v,u}, \,~ \forall v, u\, (v\neq u)\right\},
    \label{eq_ospf_parameters}
\end{equation}
where $\phi_{v,u} \in [\phi_{\min}, \phi_{\max}]$ is the weight of the physical link between routers $v$ and $u$ {($v\neq u$)}, determining the path cost and governing how packets are forwarded within the AS. $\phi_{v,u}=0$ if Routers $v$ and $u$ are not directly connected.

The full set of configurable parameters is monitored for the detection of network or configuration anomalies, as given by
\begin{equation}
    \mathcal{C} \triangleq \mathcal{C}^{\mathrm{BGP}} \cup \mathcal{C}^{\mathrm{OSPF}},
    \label{eq_configurable_parameters}
\end{equation}
where all elements of $\mathcal{C}$ take non-negative integer values, and an anomaly in any element of $\mathcal{C}$ can lead to network-wide routing disruptions or failures. 
% \end{color}

% {The configuration parameters in $\mathcal{C}$ are locally maintained at individual routers. A configuration anomaly at one router can propagate its effects to neighboring routers through BGP route advertisements and OSPF path recomputation, causing anomaly symptoms to spread across the network. Detecting such configuration anomalies requires jointly observing the configuration states of multiple routers.}

% {\color{gblue} 
% As evidenced by recent nation-scale outages~\cite{2024_ATT_outage, 2022_KDDI_outage, 2020_T_mobile_outage}, such updates are a dominant trigger of configuration anomalies, as a single inadequate change can propagate across an AS within seconds and cascade beyond before operators detect it. 
In practice, the parameter configurations of sovereign functions and protocols, e.g., $\mathcal{C}$ for BGP and OSPF, are updated across the network periodically, or whenever operational events occur, e.g., routine maintenance, software upgrades, and automated policy pushed, or the topology changes~\cite{Ray_Sovereign_Functions}, to provide situational awareness of the network. 
% 
% globally within each AS by iBGP sessions and across all ASes by eBGP sessions. 
% For example, these parameter configurations are updated every ... seconds across a ... network, or whenever new ...., 
% 
% to provide situational awareness of the network. 
% \textbf{(Not confident with this part.)}
By monitoring these configurable parameters at the network entities and route paths, potential anomalies in the protocol configurations can be early detected and mitigated.

% \begin{remark}
    
    % Intelligent agents running GNNs and detecting protocol anomalies 
    The GNNs can be deployed 
    % as lightweight software modules deployed 
    at strategically located routers, e.g., RRs and gateway routers. Each monitors and collects the local protocol configuration parameters, e.g., BGP local preference $\theta_v^\mathrm{LP}$ and OSPF link weights $\phi_{u,v}$, from its connected fact nodes, enabling distributed perception of the network state. 
    % The collected configurations are then used to update the node features $\mathbf{X}$ in $\mathcal{G}$, upon which GSID performs centralized reasoning to detect configuration anomalies across the sovereign network functions. 
    This 
    % distributed perception and centralized reasoning paradigm 
    enables scalable and timely anomaly detection.
    % without requiring every network entity to run a full inference module.}
% \end{remark}

% \red{WEI COMMENT: Add a paragraph, where do we collect the parameters? where do we execute the actions of anomaly detection? such as a router or a server?}

% To enable each protocol configuration parameter to be accurately identified, we first represent the network as a bipartite graph, which is given by
\subsection{Network Graph Representation}\label{section_graph_representation}
% \gblue{To enhance the network resilience by releasing the burden for the whole detection-recovery process, we aim to detect anomalies at the level of individual protocol configuration parameters. This is because multiple routing protocols coexist, each governing different aspects of network behavior with distinct parameters in a large-scale wireless network. We represent the network as a bipartite graph, where physical network entities (e.g., routers and ASes) and logical protocol states (e.g., BGP sessions and OSPF link configurations) form two disjoint node sets, respectively, as shown in Fig.~\ref{fig_bipartite_graph_representation}. In addition to modeling the physical entities, a bipartite graph allows protocol configuration parameters to be represented as dedicated fact nodes, enabling the anomaly detection algorithm to detect anomalies at the granularity of individual parameters across heterogeneous protocols within a unified representation.} The bipartite graph is formulated as
% As shown in Fig.~\ref{fig_bipartite_graph_representation}, 
% \begin{color}{gblue}
Inspired by \cite{NeurIPS_BGP}, we represent the considered network as a bipartite graph, $\mathcal{G} = (\mathcal{V}, \mathcal{E})$, to capture both physical topology and logical protocol states; see Fig.~\ref{fig_bipartite_graph_representation}. This representation embeds the configuration parameters into fact nodes and protocol relationships as edges. Here, $\mathcal{V}$ and $\mathcal{E}$ are the sets of vertices and edges, respectively.

% The physical network entities (e.g., routers and ASes) and logical protocol states (e.g., BGP route and OSPF link connections) form two disjoint node sets.
% \end{color}

\subsubsection{Node Representations}
% As shown in Fig.~\ref{fig_bipartite_graph_representation}, t
The node set $\mathcal{V}$ forms a bipartite structure comprising two disjoint subsets, i.e.,
% \begin{equation}
    $\mathcal{V}=\mathcal{V}_\mathrm{e} \cup \mathcal{V}_\mathrm{f}$,
% \end{equation}
where $\mathcal{V}_\mathrm{e}$ and $\mathcal{V}_\mathrm{f}$ collect entity and fact nodes, respectively. 

\begin{figure}[t]
    \centering
    \includegraphics[width=0.99\linewidth]{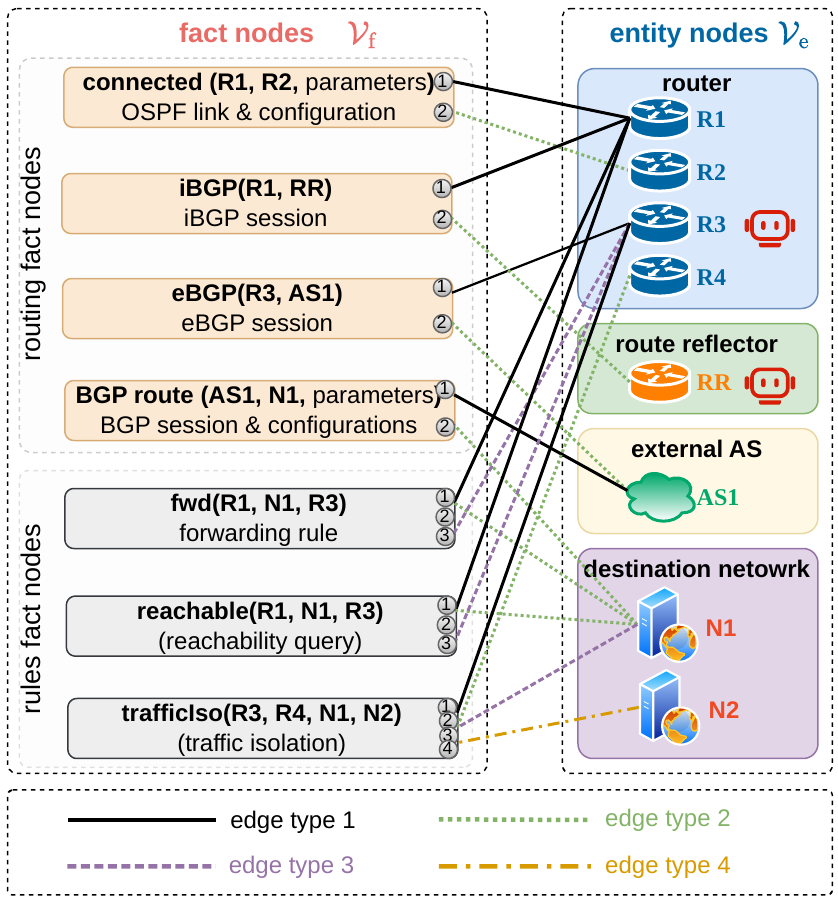}
    \caption{The bipartite graph $\mathcal{G}$ represents the physical network components and logical protocol states into two distinct node sets, denoted by entity nodes $v_\mathrm{e} \in \mathcal{V}_\mathrm{e}$ and fact nodes $v_\mathrm{f} \in \mathcal{V}_\mathrm{f}$ of the graph, respectively. By using this representation, it releases the burden for the subsequent learning algorithm for detecting anomalies across heterogeneous protocols in the network.}
    \label{fig_bipartite_graph_representation}
\end{figure}

\paragraph{Entity Nodes}
As shown in Fig.~\ref{fig_bipartite_graph_representation}, the entity node set $\mathcal{V}_\mathrm{e}$ comprises four types of physical network entities, i.e., routers, RRs, external ASes, and destination networks, consistent with that depicted in Fig.~\ref{fig_system_model}.

\paragraph{Fact Nodes}
Each fact node, $v_{\mathrm{f}}$, in $\mathcal{V}_\mathrm{f}$ corresponds to one grounded predicate instance drawn from the fact base:
\begin{equation}
v_{\mathrm{f}} \in
    \left\{
    \begin{array}{l}
    \texttt{connected}(\cdot), \\
    \texttt{iBGP}(\cdot), \\
    \texttt{eBGP}(\cdot), \\
    \texttt{BGP\_route}(\cdot), \\
    \texttt{fwd}(\cdot), \\
    \texttt{reachable}(\cdot), \\
    \texttt{trafficIso}(\cdot),
    \end{array}
    \right\}, 
\end{equation}
% {\color{red}
where $\texttt{connected}(\cdot)$, $\texttt{iBGP}(\cdot)$, $\texttt{eBGP}(\cdot)$, and $\texttt{BGP\_route}(\cdot)$, known as routing fact nodes, describe the physical link between two routers; the iBGP session between a router and an RR; the eBGP session between a router and another router in the neighboring AS; and the BGP route advertised from an external AS to its destination network, respectively.
% \red{While $\texttt{iBGP}(\cdot)$ and $\texttt{eBGP}(\cdot)$ are set up one-off at deployment or network upgrade}, 
In particular, $\texttt{connected}(\cdot)$ and $\texttt{BGP\_route}(\cdot)$ are configurable during network operations through the configurable parameters specified in \eqref{eq_bgp_parameters} and prone to configuration anomalies or inadequate configurations~\cite{SIGCOMM_Understand_BGP_Misconfig}.
% \red{(while $\texttt{iBGP}(\cdot)$ and $\texttt{eBGP}(\cdot)$ are established according to these configurable parameters.)}   
By contrast, $\texttt{fwd}(\cdot)$, $\texttt{reachable}(\cdot)$, and $\texttt{trafficIso}(\cdot)$, known as rules fact nodes, describe network intents, i.e., forwarding rules, mutual reachability/accessibility, and mutual exclusion/isolation, respectively. They 
% \red{often} 
remain unchanged during network operations. 
% By connecting entity nodes to fact nodes, we represent the ... relationship of multiple entity nodes and the choices of protocols and parameter configures substantiating or characterizing the relationship. 
% For instance, $\texttt{connected}(R1, R2, \phi_{v,u})$ encodes an OSPF link connecting routers $R1$ and $R2$ with link weight $\phi_{v,u}$; whilst $\texttt{fwd}(R1,N1,R3)$ encodes that $R1$ forwards traffic destined for $N1$ via $R3$, capturing the routing decision jointly determined by the BGP configuration parameters of $R1$. Appendix~\ref{appendix_predicate_structure} gives a complete description of all fact node types.

\begin{remark}
The bipartite graph representation $\mathcal{G} = (\mathcal{V}, \mathcal{E})$ is not restricted to BGP or OSPF. Any sovereign network function can be incorporated by defining the corresponding predicate types for $\mathcal{V}_f$ and the associated edge-type assignments $\tau \in \mathcal{T}$. The ACE and IDA modules operate directly on the resulting graph structure and require no protocol-specific modification, enabling GSID to operate in heterogeneous multi-protocol scenarios under a unified representation without protocol-specific training.
\end{remark}

\begin{figure}[t]
    \centering
    \includegraphics[width=0.999\linewidth]{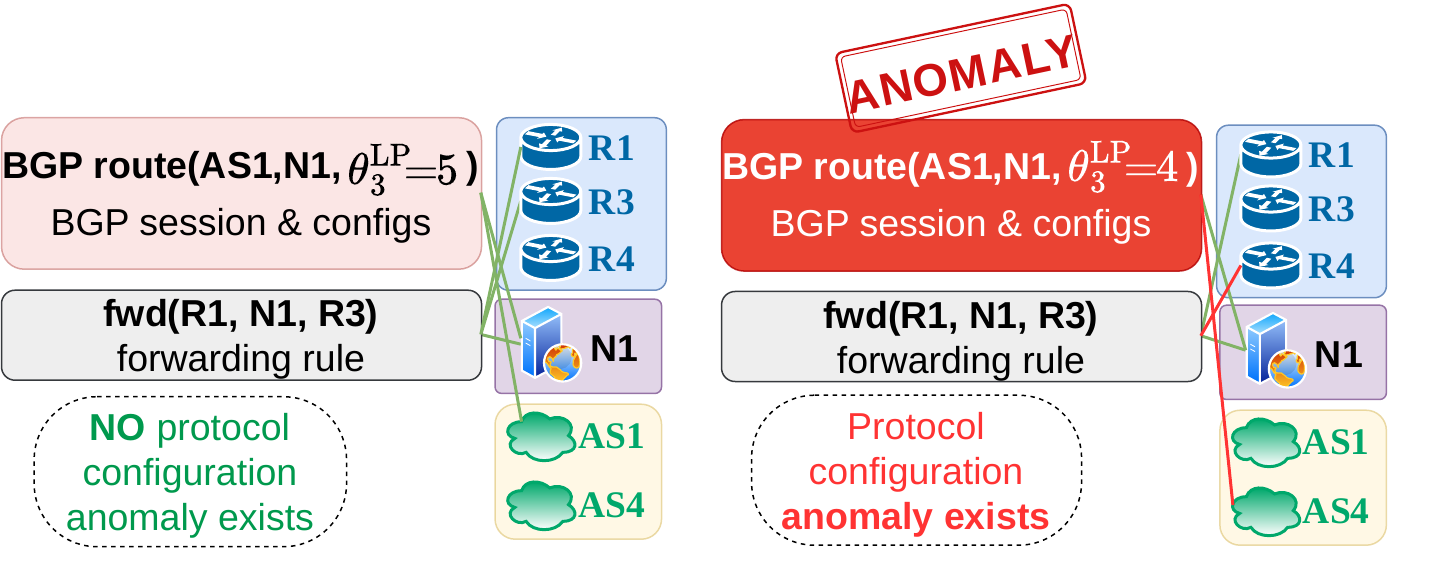}
    \caption{Illustration of protocol configuration anomaly in the bipartite graph. On the left, the \texttt{BGP\_route} fact node is correctly configured, remaining consistent with the rules fact node \texttt{fwd(R1, N1, R3)}, and traffic from R1 is forwarded to destination network N1 via the intended router R3. On the right, a configuration anomaly is injected into the BGP route fact node, introducing a semantic inconsistency with \texttt{fwd(R1, N1, R3)}. Although the misconfigured value remains valid, it redirects traffic to the unintended router R4, violating the prescribed forwarding intent. 
    {Although the misconfigured value remains valid, it redirects traffic to the unintended router R4, violating the prescribed forwarding intent. For clarity, a single-parameter anomaly is illustrated; GSID can handle multi-parameter anomaly cases via independent binary classifiers at the readout stage.}
    }
    \label{fig_anomaly}
\end{figure}

By connecting the entity and fact nodes, we characterize the relationships among network entities and the associated protocol configurations in the bipartite graph $\mathcal{G}$; see Appendix~\ref{appendix_predicate_structure} for the detailed description of all fact node types. An anomaly can be manifested as a structural inconsistency between a route involving network entities (i.e., the set of entity nodes connecting to the same routing fact node) and the protocol configuration describing the route (i.e., the feature of the routing fact node), as well as configuration rules specified by the rules fact nodes connected by the entity nodes; see Fig.~\ref{fig_anomaly}.

% a change in the routing decisions of the network, causing fact nodes to appear or disappear in $\mathcal{G}$. As illustrated in the following example, which represents the real-world configuration anomaly Example~\ref{example_configuration_anomaly_model} in the bipartite graph $\mathcal{G}$, making anomalies detectable as anomalous fact node features.

\begin{example}\label{example_graph_model} 
Fact node $\texttt{fwd}(R1, N1, R3)$ encodes that router $R1$ forwards traffic destined for destination network $N1$ through router $R3$. This forwarding is jointly determined by the BGP configuration parameters given in~\eqref{eq_bgp_parameters}. If the local preference $\theta_1^{\mathrm{LP}}$ of $R1$ is misconfigured from 5 to 4, $R1$ may select from $R3$ to a different exit point $R4$, violating the forwarding state required in the rules fact node $\texttt{fwd}(R1, N1, R3)$. A structural and semantic inconsistency arises between the route specified (i.e., from R1 to N1 via R3) and the protocol configuration describing the route. 
% In other words, unexpected or unintended edge or fact nodes, e.g. $\texttt{fwd}(R1, N1, R4)$, emerge.
% that should exist to disappear, whilst a new forwarding state $\texttt{fwd}(R1, N1, R4)$ that should not exist appears. 
\end{example}

{Notably, protocol configuration anomalies may involve multiple parameters at the same time. Since the readout stage of GSID employs independent binary classifiers for each configuration parameter, the detection across different parameters is inherently decoupled, enabling simultaneous identification of co-occurring anomalies without architectural modification.}

% \begin{remark}
% This bipartite graph representation is particularly suitable for anomaly detection, as protocol parameters are embedded in fact nodes and all information flow must pass through them, making anomalies detectable as structural inconsistencies by message passing in GNNs.
% \end{remark}

% =============================================================
% =============================================================

% =============================================================
% =============================================================

\paragraph{Node Features}
Each node in $\mathcal{V}$ is associated with a feature vector, as given by 
\begin{equation}
\begin{split}
    \mathbf{x}_v 
    & = \left[\mathbf{x}_v^{\mathrm{con}}, \mathbf{x}_v^{\mathrm{other}}\right]^\top
    \\
    & = \left[
    \theta^{\mathrm{LP}}_v,~
    \theta^{\mathrm{ASL}}_{v,u},~
    \theta^{\mathrm{MED}}_{v,u},~
    \phi_{v,u},~
    \mathbf{x}_v^{\mathrm{other}}
    \right]^\top 
    \in \mathbb{R}^{D},
    \label{eq_raw_node_feature}
\end{split}
\end{equation}
where $\mathbf{x}_v^{\mathrm{con}}=[\theta^{\mathrm{LP}}_v,~\theta^{\mathrm{ASL}}_{v,u},~\theta^{\mathrm{MED}}_{v,u},~\phi_{v,u}] \in \mathbb{R}^{|\mathcal{C}|}$ carries the configurable parameters defined in Section~\ref{section_protocol_configuration_model}, with dimensions that do not apply to a given node type set to $-1$, and $\mathbf{x}_v^{\mathrm{other}} \in \mathbb{R}^{D - |\mathcal{C}|}$ encodes other node attributes, such as node type and predicate type, with details in Appendix~\ref{appendix_feature_vector}. The feature matrix for all nodes in this network is defined as
\begin{equation}
    \mathbf{X} \triangleq \left[\mathbf{x}_{1},~ \mathbf{x}_{2},~ 
    \cdots,~ \mathbf{x}_{|\mathcal{V}|}\right]^\top \in \mathbb{R}^{|\mathcal{V}| \times D},
    \label{eq_feature_matrix}
\end{equation}

\subsubsection{Edge Representations}
As shown in Fig.~\ref{fig_bipartite_graph_representation}, non-directional edges connect the fact nodes with the entity nodes\footnote{While the edges are non-directional, the two ends of an edge are semantically different. One end is an entity node in $\mathcal{V}_\mathrm{e}$; the other end is a fact node in $\mathcal{V}_\mathrm{f}$. In light of this, dynamic attention is employed in the proposed GSID, 
as it captures this asymmetric influence from both ends of an edge by scoring the ends jointly from their concatenated embeddings.
% , implicitly distinguishing the two directions without requiring explicit directional edges;}
% where the attention on each edge is evaluated directionally from each end of the edge; 
see Section~\ref{section_design_GATv2}. From the fact node towards the entity node, the implication of rules and requirements on the compliance of routes is captured. From the entity node towards the fact node, the connectivity of entities along routes is captured.}:
\begin{equation}
    \mathcal{E} \subseteq \mathcal{V}_\mathrm{e} \times \mathcal{V}_\mathrm{f}.
\end{equation}
Moreover, $\mathcal{V}_\mathrm{e}$ and $\mathcal{V}_\mathrm{f}$ are disjoint, so no edges are permitted within either of the sets.
Each fact node (left column of Fig.~\ref{fig_bipartite_graph_representation}) can be linked to multiple entity nodes (right column of Fig.~\ref{fig_bipartite_graph_representation}) that participate in the corresponding protocol execution. Each edge is assigned an edge type $\tau$ that is encoded by the position of the connected entity in the fact node, 
\begin{equation}
    \tau(u,v) \in \mathcal{T}, \quad \forall (u,v) \in \mathcal{E}.
\end{equation}
Every fact node is connected to at least two entity nodes, where $\mathcal{N}(v) \triangleq \{u \in \mathcal{V}_\mathrm{e} \mid (u,v) \in \mathcal{E}\}$ denotes the neighbor set of $v$.
\begin{equation}
    |\mathcal{N}(v)| \geq 2, \quad \forall v \in \mathcal{V}_\mathrm{f}.
\end{equation}

\begin{example}
$\texttt{iBGP}(c_i, c_j)$ generates two edges, with an edge of $\tau{=}0$ connecting router $c_i$, and an edge of $\tau{=}1$ connecting RR $c_j$. Similarly, $\texttt{fwd}(c_n, c_m, c_k)$ generates three edges, with $\tau{=}0$ connecting source router $c_n$, $\tau{=}1$ connecting destination network $c_m$, and $\tau{=}2$ connecting next-hop router $c_k$. 
The correct connection between the routing fact node describing a route, network entities to be involved in the route, and the rules fact nodes concerning the network entities represents a legitimate protocol configuration.
Please refer to Table~\ref{tab_fact_node_args} in Appendix~\ref{appendix_predicate_structure} for the edge-type assignments of all fact nodes.
\end{example}

% With this indicator, we formulate the considered protocol configuration anomaly detection problem as learning a mapping, as given by
% \begin{equation}
%     f_{\boldsymbol{\omega}} : \left(\mathcal{G},\,\tilde{\mathbf{X}}\right) 
%     \mapsto \hat{\mathbf{Y}},
%     \label{eq_problem_formulation}
% \end{equation}
% where $\boldsymbol{\omega}$ denotes the learnable parameters, $\tilde{\mathbf{X}} \in \mathbb{R}^{|\mathcal{V}|\times D}$ is the observed feature matrix over all nodes, and $\hat{\mathbf{Y}} \in \{0,1\}^{|\mathcal{V}_\mathrm{f}|\times|\mathcal{K}|}$ is the predicted anomaly label matrix comprising the anomaly indicators $y_{v,k},\,\forall v,k$.
% 

\subsection{Configuration Anomaly}\label{section_anomaly_model}

An anomaly is an error that sets one or more protocol parameters inappropriately, causing the network to malfunction, e.g., forwarding traffic along unintended paths, even without necessarily triggering an obvious failure or network outage. 
% 
% Following~\cite{NeurIPS_BGP}, we monitor four configurable protocol parameters of BGP and OSPF protocols in this work. 
% For BGP, we monitor three parameters: i) local preference set within the AS, specifying the preferred exit point when routing traffic to external destinations. Higher values indicate higher priority. ii) AS path length used to calculate the number of ASes traversed by a specific route. Shorter paths indicate higher priority. iii) Multi-exit discriminator (MED) used to advertise to neighboring ASes, suggesting the preferred entry point into the local AS. Lower values indicate higher priority. 
% For OSPF, we monitor one parameter, the link weight, which determines the cost of each physical link in Dijkstra's shortest path calculation. 
% Collectively, these four parameters control the routing of the traffic data; an anomaly in any one of these parameters can lead to network-wide routing disruptions. For more detailed information on the BGP and OSPF protocols, please refer to Appendix~\ref{section_appendix_BGP_OSPF}.
% 

\begin{example}\label{example_configuration_anomaly_model}
When the local preference parameter $\theta_{1}^{\mathrm{LP}}$ of router $R1$ is unintentionally configured from $5$ (i.e., preferring $R3$ as the exit point) to $4$ (i.e., preferring $R4$ as the exit point), {leading to traffic} being forwarded along an unintended path across the AS without triggering any explicit fault signal.
Such an anomaly can be caused by human-induced error during manual configuration or automated policy updates that fail to account for the current network topology~\cite{SIGCOMM_Understand_BGP_Misconfig}.
% it is possible that R1 selects a suboptimal exit point, causing traffic to be forwarded along an unintended path across the AS. Such configuration anomaly can be caused by human-induced error during manual configuration or automated policy updates that fail to account for the current network topology.
\end{example}

We define a configuration anomaly as
\begin{equation}
\label{eq_configuration_anomaly_defintiion}
    \hat{x}_{v,c}\! \neq\! x_{v,c}^*,  \forall \hat{x}_{v,c} \!\in \!\!
    \left\{\begin{array}{@{}l@{}l@{}}
        [\theta^{\mathrm{LP}}_{\min},  \theta^{\mathrm{LP}}_{\max}],  & \text{ if } \hat{x}_{v,c} \!= \!\theta_{v,c}^{\mathrm{LP}};  \\[8pt]
        [\theta^{\mathrm{ASL}}_{\min}, \theta^{\mathrm{ASL}}_{\max}], & \text{ if } \hat{x}_{v,c}\! =\! \theta_{v,c}^{\mathrm{ASL}}; \\[8pt]
        [\theta^{\mathrm{MED}}_{\min}, \theta^{\mathrm{MED}}_{\max}], & \text{ if } \hat{x}_{v,c} \!= \!\theta_{v,c}^{\mathrm{MED}}; \\[8pt]
        [\phi_{\min}, \phi_{\max}],                                   & \text{ if } \hat{x}_{v,c} \!= \!\phi_{v,c},
    \end{array}\right.
\end{equation}
where $\hat{x}_{v,c} \in \hat{\mathbf{x}}_v$ is the observation of the $c$-th protocol configuration parameter of node $v$; and $x_{v,c}^* \in \mathbf{x}_v^*$ is the corresponding ground truth that suits the routing policy and design intent, e.g., intended forwarding paths, traffic isolation, and reachability. 

% {\color{red} \textbf{COMMENT:} Please explain (4) and clarify its physical meaning.}

The four cases in~\eqref{eq_configuration_anomaly_defintiion} correspond to the four protocol parameters described in Section~\ref{section_protocol_configuration_model}, where the potentially anomalous value of each parameter (e.g., $\hat{x}_{v,c}$) remains within its valid range, making the anomaly semantically subtle, since it cannot be identified by inspecting individual parameters in isolation, but only by assessing the collective consistency among parameter configurations across the network and their compliance with the preset networking intents. 

The difficulty of detection is exacerbated by the fact that each parameter, despite appearing legitimate individually, carries distinct routing consequences that can cascade across the network. Specifically, 
i) an anomalous local preference $\theta_{v}^{\mathrm{LP}} \in [\theta^{\mathrm{LP}}_{\min}, \theta^{\mathrm{LP}}_{\max}]$ can redirect the exit point selection across all iBGP sessions connected to the $v$-th router; 
ii) an anomalous AS path length $\theta_{v,u}^{\mathrm{ASL}} \in [\theta^{\mathrm{ASL}}_{\min}, \theta^{\mathrm{ASL}}_{\max}]$ manipulates route preference by altering the perceived path length of the route between routers $v$ and $u$; 
iii) an anomalous MED $\theta_{v,u}^{\mathrm{MED}} \in [\theta^{\mathrm{MED}}_{\min}, \theta^{\mathrm{MED}}_{\max}]$ misleads neighboring ASes to selecting an unintended entry point into the $v$-th router's AS, diverting inter-AS traffic through an unintended ingress;
iv) an anomalous OSPF link weight $\phi_{v,u} \in [\phi_{\min}, \phi_{\max}]$ corrupts the intra-AS shortest path computation between routers $v$ and $u$, causing traffic to be forwarded along unintended paths.

\begin{remark}\label{remark_anomaly}
    An anomalous configuration value of the protocols can still remain within the protocol-defined valid range and thus does not raise an explicit fault signal, making detection non-trivial.
    % The anomaly in even a single protocol parameter can propagate across the entire network and trigger cascading failures in large-scale networks. 
    For example, given that both $\theta^{\mathrm{LP}}_1 = 4$ and $\theta^{\mathrm{LP}}_1 = 5$ lie within $[\theta^{\mathrm{LP}}_{\min}, \theta^{\mathrm{LP}}_{\max}]$, a configuration anomaly from $5$ to $4$ raises no fault signal and is indistinguishable from a legitimate configuration.
    But the consequence would be redirecting traffic away from the intended exit point R3 to R4, propagating across the AS without any explicit fault signal; see Example~\ref{example_graph_model}.
    % an inappropriate BGP local preference value can cascade across all iBGP sessions that redistribute the affected route, potentially changing the exit-point selection of every router within the AS. 
    %Similarly, a misconfigured OSPF link weight can silently reroute traffic along suboptimal paths by altering the cost metric in Dijkstra's computation across the entire AS.
\end{remark}

\subsection{Problem Formulation}
As defined in~\eqref{eq_raw_node_feature}, each fact node $v_\mathrm{f} \in \mathcal{V}_\mathrm{f}$ carries a feature vector $\mathbf{x}_v$, with, $\hat{x}_{v,c} \in \hat{\mathbf{x}}_v$, the observed value of the $c$-th configuration parameter of node $v$ with potential anomalies, and, $x_{v,c}^* \in \mathbf{x}_v^*$, the corresponding ground-truth configuration value, the ground-truth anomaly label is given by
\begin{equation}
    y_{v,c}^* = \mathds{1}\{\hat{x}_{v,c} \neq x_{v,c}^*\},
    \label{eq_anomaly_label}
\end{equation}
where $\mathds{1}\{\cdot\}$ is the indicator function, indicating $y_{v,c}^*=1$ if the observed value $\hat{x}_{v,c}$ deviates from the ground-truth configuration value $x_{v,c}^*$, and 0 otherwise. 

Given the bipartite graph $\mathcal{G} = (\mathcal{V}, \mathcal{E})$ 
% in~\eqref{eq_graph_definition}
and the observed feature matrix $\hat{\mathbf{X}}$ with potential configuration anomalies (e.g., Example~\ref{example_graph_model}), the goal of our design is to learn the mapping
\begin{equation}
    \tilde{\mathbf{Y}} 
    = f_{\boldsymbol{\omega}} \!\left(\mathcal{G},~\hat{\mathbf{X}}\right) 
    \in [0,1]^{|\mathcal{V}_\mathrm{f}| \times |\mathcal{C}|},
\end{equation}
where $\tilde{\mathbf{Y}}$ is the predicted anomaly matrix. Each entry $\tilde{y}_{v,c} = [\tilde{\mathbf{Y}}]_{v,c} \in \{0,1\}$ represents the predicated anomaly indicator of the $c$-th configuration parameter at fact node $v$. 

With the ground-truth anomaly labels $y_{v,c}^*$ available, the problem is formulated as a supervised learning problem, 
\begin{equation}
    \boldsymbol{\omega}^*=\arg \min_{\boldsymbol{\omega}} \sum_{v \in \mathcal{V}_\mathrm{f}} 
    \sum_{c \in \mathcal{C}} 
    \mathcal{L}\!\left(\tilde{y}_{v,c},~ y_{v,c}^*\right),
\end{equation}
where the optimal parameters $\boldsymbol{\omega}^*$ are obtained by minimizing the cross-entropy loss $\mathcal{L}(\cdot, \cdot)$.

\begin{figure*}[t]
    \centering
    \includegraphics[width=0.99\linewidth]{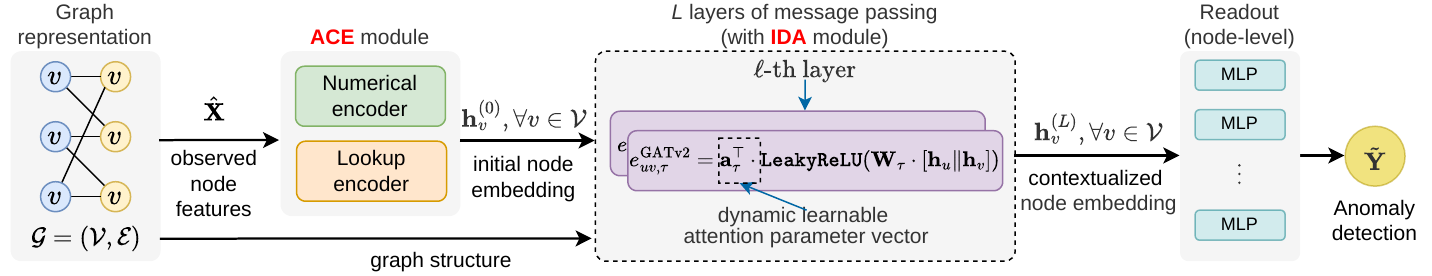}
    \caption{Overall architecture of the proposed GSID, which first encodes raw node features using the CA node feature encoder, then performs $L$ layers of topology-aware message passing to capture network-wide dependencies, and lastly employs a set of node-level classifiers for network anomalies caused by configuration inconsistency, configuration anomaly, etc.}
    \label{fig_architecture}
\end{figure*}
% ====================================================================

\section{Proposed GSID}\label{section_proposed_algorithm}
In this section, we introduce the proposed GSID, and delineate its two core designs, i.e., the CA node feature encoder and the IDA mechanism, followed by the computational complexity analysis of GSID. % is analyzed.
% The two designs are motivated by the role of GSID as a context-driven perception module within an agentic AI pipeline for network resilience: effective agentic decision-making over dynamic wireless networks requires an AI that can exploit the rich contextual signals provided by the network --- including heterogeneous protocol configurations and non-uniform anomaly propagation patterns --- to accurately diagnose faults before any remediation action can be taken. The configuration encoder and IDA mechanism are specifically designed to extract and leverage these two forms of network context, respectively. 
\subsection{Design of GSID Architecture}
As shown in Fig.~\ref{fig_architecture}, our GSID processes the input $\mathcal{G}$ through two new modules followed by a standard node-level readout. The first module is the proposed CA node feature encoder, which maps the observed node features $\hat{\mathbf{X}}$ into initial embeddings $\mathbf{h}_v^{(0)}, \forall v \in \mathcal{V}$ by adaptively selecting encoding strategies for each configuration parameter. 
% This design addresses the heterogeneity of protocol configuration parameters \red{identified in Challenge 1 in the Introduction}. 
% 
The second module is the IDA-guided $L$-layer message-passing, which produces the contextualized node embedding $\mathbf{h}_v^{(L)}, \forall v \in \mathcal{V}$ by aggregating neighborhood information with dynamic attention. The IDA module scores each edge after the nonlinear activation function on the concatenated embeddings of the two ends. This dynamic attention mechanism aggregates information from both ends of each edge to exploit the asymmetric influence
% Attention is drawn from both ends to account for the different semantics 
of fact and entity nodes in the bipartite graph $\mathcal{G}$.
% caused by the directional non-linearity of the anomalies, which is reflected as the different edge directions from fact node to entity node and from entity node to fact node.}
% since an anomaly can be viewed as an inconsistency between the set of edges connecting a fact node and the protocol configuration described by the fact node.
% This design addresses the challenge of \red{spatially localized effects} caused by configuration anomalies, as described in Section~\ref{section_introduction}.
% % 
% capturing whether anomaly symptoms propagate between two nodes based on their joint configuration state. This design addresses Challenge 2 in the Introduction.
Lastly, the standard node-level readout module comprising $|\mathcal{C}|$ independent multilayer perceptrons (MLPs) produces a binary anomaly prediction for each monitored configuration feature.

\subsection{Design of Adaptive Configuration Encoder}\label{section_design_encoder}
The ACE is designed to address the heterogeneous nature of protocol configuration parameters. As described in~\eqref{eq_raw_node_feature}, BGP determines the preferred exit point by enabling each router to compare the configurable parameters (i.e., $\theta_{v}^{\mathrm{LP}}$, $\theta_{v,u}^{\mathrm{ASL}}$, and $\theta_{v,u}^{\mathrm{MED}}$) in order of priority; OSPF selects the shortest path by minimizing the cumulative link weight \(\phi_{v,u}\) using Dijkstra's algorithm~\cite{Dijkstra}, and $\mathbf{x}^\mathrm{other}$ encodes categorical parameters, such as node types. As a result, $\mathbf{x}^\mathrm{con}=[\theta^{\mathrm{LP}}_v,~\theta^{\mathrm{ASL}}_v,~\theta^{\mathrm{MED}}_v,~\phi_{v,u}]$ governs routing decisions through direct comparison, whilst $\mathbf{x}^\mathrm{other}$ encodes discrete protocol identities for resource-efficient pattern recognition.

\begin{example}
    A local preference $\theta_{v}^\mathrm{LP}$ deviated from 5 to 4 may unnoticeably cascade route selection changes across all iBGP routes associated with the $v$-th router, without raising any immediate outages. An ideal encoder is expected to preserve how far the value has deviated. In contrast, a node type parameter encodes a discrete identity, e.g., router or RR, so a lookup encoder is therefore adopted to distinguish the type in a resource-efficient manner.
\end{example}

% ====================================================================
\begin{figure}[t]
    \centering
    \includegraphics[width=0.99\linewidth]{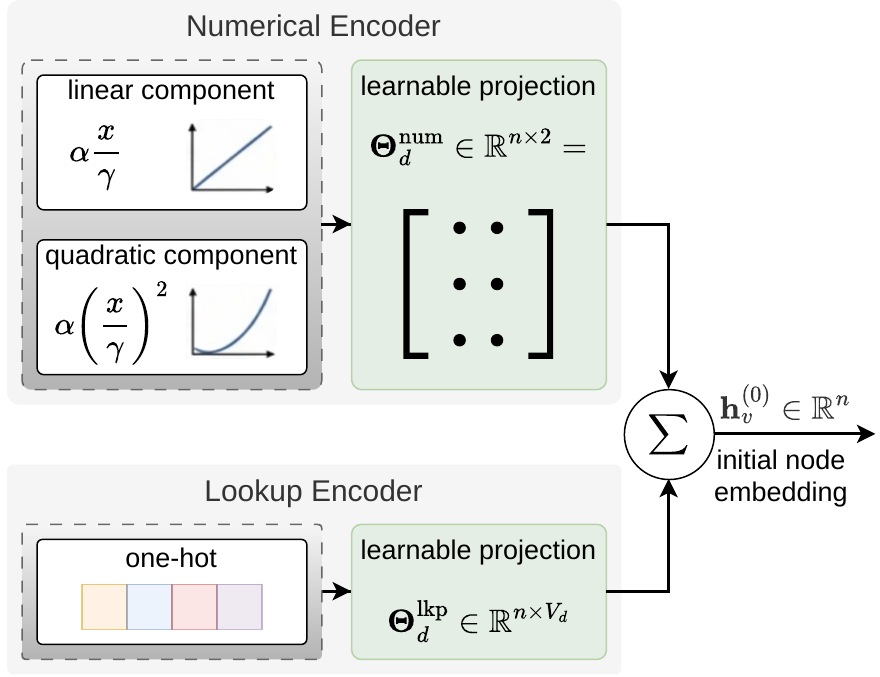}
    \caption{Architecture of the ACE. Numerical protocol parameters are mapped through a dual-path nonlinear encoder to preserve adaptive structure and enhance sensitivity to large perturbations, while categorical attributes are embedded via a standard lookup mechanism.}
    \label{fig_CA_encoder}
\end{figure}
% ====================================================================

As shown in Fig.~\ref {fig_CA_encoder}, the ACE selects a numerical encoder or a lookup encoder as a sub-encoder for each feature according to its configuration type, and accumulates the results to produce the initial node embedding as
\begin{equation}
\begin{split}
    \mathbf{h}_v^{(0)} 
    &= f^\mathrm{ACE}(\hat{\mathbf{x}}_v)
    = \sum_{d=1}^{D} \Big(\mathds{1}\{\hat{x}_{v,d} \neq -1\} \cdot f_d^\mathrm{en} \!\left(\hat{x}_{v,d}\right) \Big),
\label{eq_configuration encoder_encoder}
\end{split}
\end{equation}
where $\hat{x}_{v,d}$ is the $d$-th element of the observed feature vector of the $v$-th node, $\hat{\mathbf{x}}_v$, and the indicator $\mathds{1}\{x_{v,d} \neq -1\}$ excludes inapplicable dimensions (see Appendix~\ref{appendix_feature_vector} for details). Function $f_d^\mathrm{en}(\cdot)$ represents the $d$-th sub-encoder selected, as given by
\begin{equation}
    f_d^\mathrm{en}(\hat{x}_{v,d}) =
    \begin{cases}
        f^\mathrm{num} \left(\hat{x}_{v,d} \right), & \text{if } d \in \mathcal{M}; \\
        f^\mathrm{lkp} \left(\hat{x}_{v,d} \right), & \text{otherwise},
    \end{cases}
    \label{eq_sub_encoder_selection}
\end{equation}
where $f^\mathrm{num}(\cdot)$ and $f^\mathrm{lkp}(\cdot)$ denote the numerical encoder and the lookup encoder, respectively; and $\mathcal{M}$ denotes the index set of numerical features, comprising BGP local preference, AS path length, MED, and OSPF link weight.

\subsubsection{Numerical Encoder}
Given a scalar numerical feature $x$, the numerical encoder computes a two-dimensional nonlinear feature map and projects it to the hidden dimension $n$ as
\begin{equation}
    f_d^\mathrm{num}(\hat{x})
    = \mathbf{\Theta}_d^\mathrm{num} \cdot
      \left[
          \varepsilon_1 \frac{\hat{x}}{\gamma},\;
          \varepsilon_2 \left(\frac{\hat{x}}{\gamma}\right)^{\! 2}
      \right]^{\top}
    \in \mathbb{R}^{n},
    \label{eq_numerical_encoder}
\end{equation}
where $\gamma > 0$ is a normalization constant; $\varepsilon_1, \varepsilon_2 > 0$ are scaling coefficients for the linear and quadratic components, respectively; and $\mathbf{\Theta}_d^\mathrm{num} \in \mathbb{R}^{n \times 2}$ is a learnable projection matrix specific to the features in the $d$-th dimension. 

The linear component preserves the adaptive magnitude of the parameter, ensuring that numerically proximate values produce similar embeddings. The quadratic component provides asymmetric sensitivity that small deviations near zero produce modest embedding differences, while larger deviations are amplified, reflecting the observation that larger anomaly magnitudes tend to produce more pronounced network symptoms. 
{In real-world scenarios, $\gamma$ can be configured according to the protocol-defined valid range of each parameter (e.g., $[\theta^{\mathrm{LP}}_{\min}, \theta^{\mathrm{LP}}_{\max}]$), so that the normalized input $\hat{x}/\gamma$ remains in a bounded range irrespective of the absolute scale of the configuration parameter, preventing large input magnitudes from producing excessively large embedding values. 
The scaling coefficients $\varepsilon_1, \varepsilon_2 > 0$ govern the parameter sensitivity of the numerical encoder: a larger $\varepsilon_2$ relative to $\varepsilon_1$ increases the sensitivity to larger deviations while keeping the response to small perturbations modest, thereby providing asymmetric anomaly amplification without causing gradient instability during training.}

\subsubsection{Lookup Encoder}
For the remaining $D - |\mathcal{M}|$ features, the values of the configuration parameters represent discrete identities, which can be encoded by a resource-efficient lookup encoder. The lookup sub-encoder is designed as 
\begin{equation}
    f_d^\mathrm{lkp}(x)
    = \mathbf{\Theta}_d^\mathrm{lkp} \cdot \boldsymbol{\delta}_x \in \mathbb{R}^{n},
    \label{eq_lookup_encoder}
\end{equation}
where $\mathbf{\Theta}_d^\mathrm{lkp} \in \mathbb{R}^{n \times V_d}$ is a learnable projection matrix specific to the $d$-th node feature, and $\boldsymbol{\delta}_x \in \{0,1\}^{V_d}$ is the one-hot representation vector of $x$, indicating the discrete identity of the feature value.

% ====================================================================
\begin{figure}[t]
    \centering
    \includegraphics[width=0.99\linewidth]{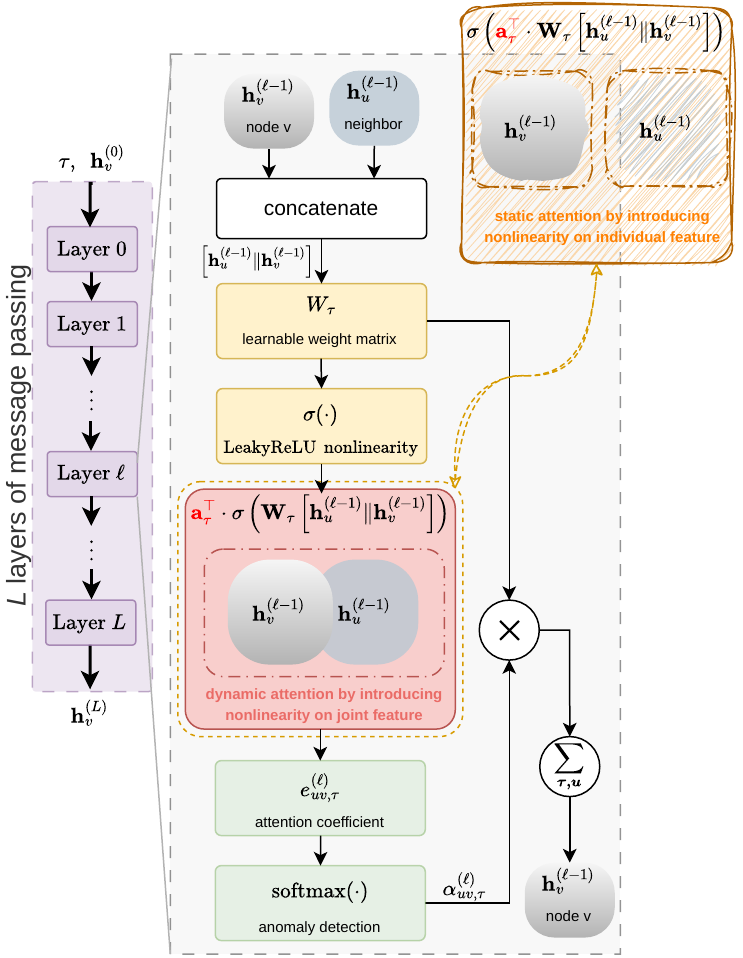}
    \caption{
    % \red{\textbf{Need help}. This figure lacks telecom information. This figure looks like a pure AI module, worrying that it would be challenged by reviewers for lacking novelty by just applying GATv2.} 
    Architecture of the IDA mechanism. The attention vector $\mathbf{a}$ is applied after the non-linearity to ensure a dynamic scoring from both ends of each edge. As it captures the asymmetric influence of fact nodes and entity nodes, reflected by the different semantics from the two ends of edges in the bipartite graph. Information aggregation is guided by transformation matrices $\mathbf{W}_\tau$ to respect the distinct semantics of different communication protocols.}
    \label{fig_dynamic_attention}
\end{figure}

\subsection{Design of Inconsistency Dynamic Attention}\label{section_design_GATv2}
As shown in Fig.~\ref{fig_dynamic_attention}, given the embedded node features $\mathbf{h}_v^{(0)} \in \mathbb{R}^n$ from \eqref{eq_configuration encoder_encoder}, where $n$ is the hidden embedding dimension shared across all layers, and GSID performs $L$ layers of message passing over the graph $\mathcal{G}$. At the $\ell$-th layer, we compute the dynamic attention coefficient\footnote{The attention coefficient in GAT~\cite{GAT_ICLR_2018} is $e_{uv}^\mathrm{GAT} = \texttt{LeakyReLU}(\mathbf{a}^\top \cdot  {\mathbf{W}_{\tau}} [\mathbf{h}_u \| \mathbf{h}_v])$, where $\mathbf{a}$ is placed before the non-linearity by activation function, collapsing with a learnable weight matrix ${\mathbf{W}_{\tau}}$.} between nodes $u$ and $v$ connected by an edge of type $\tau$, and is given by
\footnote{Dynamic attention, originally developed in~\cite{GATv2} to capture asymmetric influence of both ends on an edge, is suited to detecting protocol configuration anomalies in the bipartite graph since protocol configuration anomalies are interpreted as structural and semantic inconsistencies between the set of edges connecting a routing fact node, the protocol configuration described by the routing fact node, and rules and requirements specified by the rules fact nodes connected with the entity nodes involved in the route{, going beyond the general-purpose dynamic attention of GATv2~\cite{GATv2} by exploiting the domain-specific bipartite structure of the communication network.}}
\begin{equation}
    e_{uv,\tau}^{(\ell)}
    = \mathbf{a}_\tau^\top \cdot
      \sigma
      \left(
        \mathbf{W}_\tau 
        \left[
        \mathbf{h}_u^{(\ell-1)}
        \,\middle\|\,
        \mathbf{h}_v^{(\ell-1)}
      \right]
      \right),
    \label{eq_gatv2_score}
\end{equation}
where $\mathbf{a}_\tau \in \mathbb{R}^{n}$ is a learnable attention vector specific to edge type $\tau$, $\sigma(\cdot)$ is the non-linear $\operatorname{LeakyReLU}$ activation function, $\mathbf{W}_\tau \in \mathbb{R}^{n \times 2n}$ is applied to the concatenated embeddings of the two endpoints, and $[\cdot\|\cdot]$ denotes concatenation. 
The coefficient is then normalized over all neighbors of type $\tau$ using $\operatorname{softmax}$ to obtain the attention weight, as given by
\begin{equation}
    \alpha_{uv,\tau}^{(\ell)}
    = \operatorname{softmax}\!\left( e_{uv,\tau}^{(\ell)} \right).
    \label{eq_attention_score}
\end{equation}
The embedding of node $v$ at layer $\ell$ is updated by aggregating weighted messages from its neighbors across all edge types:
\begin{equation}
    \mathbf{h}_v^{(\ell)}
    = \sum_{\tau \in \mathcal{T}}
      \sum_{u \in \mathcal{N}_\tau(v)}
      \alpha_{uv,\tau}^{(\ell)} \cdot \mathbf{W}_\tau\,
      \mathbf{h}_u^{(\ell-1)},
    \label{eq_CT-Ca_message_passing}
\end{equation}
where $\mathcal{N}_\tau(v)$ denotes the set of neighbors connected to node $v$ with edges of type $\tau$.

After $L$ layers of message passing, the contextualized node embedding $\mathbf{h}_v^{(L)}$ is passed to the node-level binary classifier for the detection of incorrect or unexpected configurations. For the $k$-th monitored feature of node $v$, the readout produces
\begin{equation}
    \hat{y}_{v,c}
    = \operatorname{softmax}\!\left(
        \operatorname{MLP}^{(c)}\!\left( \mathbf{h}_v^{(L)} \right)
      \right) \in \mathbb{R}^{2},
    \label{eq_readout_binary}
\end{equation}
which gives the probability that the $k$-th feature of node $v$ has been tampered with.

% ====================================================================
\begin{algorithm}[!t]
\algsetup{linenosize=\small} \small 
\SetAlgoLined
\caption{The Proposed GSID Algorithm}
\label{alg_CT-Ca}
\KwIn{Training dataset $\mathcal{D} = \{\mathcal{G}_1, \ldots, \mathcal{G}_N\}$, where each $\mathcal{G}_i = (\mathcal{V}_i, \mathcal{E}_i, \mathbf{X}_i, \mathcal{T}_i)$, with $\mathbf{X}_i \in \mathbb{R}^{|\mathcal{V}_i| \times D}$ denoting the raw node feature matrix and $\mathcal{T}_i$ the edge-type set; number of monitored features per node $|\mathcal{C}|$; learning rate $\eta$, training epochs $T_{\max}$, hidden dimension $n$, number of attention heads $H$, number of encoder layers $L$, number of decoder iterations $N_\mathrm{iter}$.}

\KwOut{Trained neural network parameters $\boldsymbol{\omega}^*$.}

\textbf{Initialize:} model parameters $\boldsymbol{\omega}$, comprising the CA node feature encoder $f^{\mathrm{ACE}}(\cdot)$ with numerical encoders $\{f_d^{\mathrm{num}}(\cdot)\}$ and lookup encoders $\{f_d^{\mathrm{lkp}}(\cdot)\}$, the GATv2 layers with parameters $\{\mathbf{W}_\tau, \mathbf{a}_\tau\}_{\tau \in \mathcal{T}}$, and per-feature binary classifiers $\{\mathrm{MLP}^{(c)}\}_{c \in \mathcal{C}}$.

\For{$t = 1, \ldots, T_{\max}$}{

  \For{each mini-batch $\mathcal{B} \subseteq \mathcal{D}$}{

    Add bidirectional edges and self-loops to each graph sample $\mathcal{G}_i \in \mathcal{B}$.\\

    Randomly inject anomalies into $\{\mathcal{G}_i\}_{i \in \mathcal{B}}$ to obtain ground-truth labels $\{y_{v,c}\}$.\\

    Encode node features into initial embeddings $\mathbf{h}_v^{(0)}$ via the CA node feature encoder using~\eqref{eq_configuration encoder_encoder}--\eqref{eq_lookup_encoder}.\\

    \For{$\ell = 1, \ldots, L$}{

        Compute attention weights $\alpha_{uv,\tau}^{(\ell)}$ per edge type $\tau$ by~\eqref{eq_attention_score}.\\

        Update node embeddings $\mathbf{h}_v^{(\ell)}$ by message passing using~\eqref{eq_CT-Ca_message_passing}.

    }

    \For{$i = 1, \ldots, N_{\rm iter}$}{

        Update node embeddings $\mathbf{h}_v$ by message passing using~\eqref{eq_CT-Ca_message_passing} (with residual connection).

    }

    Produce per-feature anomaly predictions $\{\hat{y}_{v,c}\}$ from $\mathbf{h}_v$ by~\eqref{eq_readout_binary}.\\

    Compute cross-entropy loss $\mathcal{L}\left(\{\hat{y}_{v,c}\}, \{y_{v,c}\}\right)$ and update $\boldsymbol{\omega} \leftarrow \boldsymbol{\omega} - \eta\,\nabla_{\boldsymbol{\omega}} \mathcal{L}$.

  }

}

\Return $\boldsymbol{\omega}^* = \boldsymbol{\omega}$.
\end{algorithm}
% ====================================================================

\subsection{Training Algorithm}
The training procedure is summarized in Algorithm~\ref{alg_CT-Ca}, which first encodes node features, e.g., BGP configuration parameters ($\theta_{v}^{\mathrm{LP}}$, $\theta_{v,u}^{\mathrm{ASL}}$, and $\theta_{v,u}^{\mathrm{MED}}$), OSPF link weights ($\theta_{v,u}$), and categorical node attributes like node type and predicate type detailed in Appendix~\ref{appendix_feature_vector}, into initial embeddings by the ACE. Next, $L$ rounds of message passing update node embeddings using the IDA mechanism. Lastly, the node-level binary classifiers produce anomaly predictions from the final embeddings $\mathbf{h}^{(L)}$, and the model parameters are updated by minimizing the cross-entropy loss~$\mathcal{L}(\cdot, \cdot)$.

% \footnote{\red{ (delete?) Empirical experiments demonstrated that the specific design of the numerical encoder is unlimited to \eqref{eq_numerical_encoder}, other non-linear designs can also be useful.}}. 

\subsection{Computational Complexity Analysis}\label{section_complexity}
We analyze the computational complexity of GSID in two stages, i.e., raw node feature encoding and message passing. %, and readout classifier.

\subsubsection{Adaptive Configuration Encoder}
For a graph with $|\mathcal{V}|$ nodes with each node carrying a $D$-dimensional raw feature vector, the proposed ACE in~\eqref{eq_configuration encoder_encoder} processes each feature dimension independently using the sub-encoders as numerical encoder or lookup encoder; see~\eqref{eq_sub_encoder_selection}. 

The numerical encoder in~\eqref{eq_numerical_encoder} projects a two-dimensional nonlinear feature map to the field $\mathbb{R}^{n}$ with $\mathbf{\Theta}_d^{\mathrm{num}} \in \mathbb{R}^{n \times 2}$, incurring $\mathcal{O}(n)$ operations per feature.
The lookup encoder in~\eqref{eq_lookup_encoder} retrieves the corresponding column of $\mathbf{\Theta}_d^{\mathrm{lkp}} \in \mathbb{R}^{n \times V_d}$ by index, incurring $\mathcal{O}(1)$ per feature. Considering all $D$ features of $|\mathcal{V}|$ nodes, the encoding stage incurs $\mathcal{O}\!\left(|\mathcal{V}| \cdot |\mathcal{M}| \cdot n \right)$, where $|\mathcal{M}|$ is the number of features that we choose to encode by the numerical sub-encoder.
% is fixed by the protocol feature set. The encoder scales linearly with~$|\mathcal{V}|$.

% {\color{red} \textbf{(Shall we move this paragraph to Section~\ref{section_design_encoder}?)} 
The conclusion drawn is that the more complex (in terms of $\theta_{v}^\mathrm{LP}$, $\theta_{v,u}^\mathrm{ASL}$, $\theta_{v,u}^\mathrm{MED}$, and $\phi_{v,u}$) the configuration parameters are, the higher the non-linearity is needed to predict the anomaly in it. 
Finer-grained encoding strategies, e.g., cubic-order polynomial, could be assigned to individual parameter types, e.g., $\phi_{v,u}$, to improve detection accuracy. However, this comes at the cost of increased model complexity and reduced generalizability, as a more specialized encoder requires meticulous design and may overfit to the specific parameter distributions observed during training.
% }

\subsubsection{Inconsistency Dynamic Attention-based Message Passing}
At each of the $L$ message-passing layers, GSID computes the dynamic attention coefficients and aggregates messages over all typed edges. For an edge of type $\tau$ between fact and entity nodes $u$ and $v$, the attention score in \eqref{eq_gatv2_score} requires a linear transform $\mathbf{W}_\tau \in \mathbb{R}^{n \times 2n}$ applied to the concatenated embeddings of the two nodes, followed by a dot product with $\mathbf{a}_\tau \in \mathbb{R}^{n}$, incurring a computational complexity of $\mathcal{O}(n^2)$ per edge. Aggregating over all edge types and edges, the cost is $\mathcal{O}(|\mathcal{E}| \cdot n^2)$ per layer, where $|\mathcal{E}| = \sum_{\tau \in \mathcal{T}} |\mathcal{E}_\tau|$ is the total number of typed edges. Over $L$ layers, the message-passing stage incurs a computational complexity of $\mathcal{O}\!\left(L \cdot |\mathcal{E}| \cdot n^2\right)$.
Since each node's embedding $\mathbf{h}_v^{(\ell)}$ in~\eqref{eq_CT-Ca_message_passing} is the summation of the weighted neighbor messages across all $|\mathcal{T}|$ edge types, the dependence on $|\mathcal{T}|$ is captured by the edge traversal over~$|\mathcal{E}|$.

% \subsubsection{Readout Classifier}
% The per-feature binary classifier in \eqref{eq_readout_binary} applies $K$ independent MLPs, each mapping $\mathbb{R}^{n}$ to $\mathbb{R}^2$. Let $n'$ denote the number of hidden units per MLP. The readout stage over all $|\mathcal{V}|$ nodes incurs 
% \begin{equation}
%     \mathcal{O}\!\left(|\mathcal{V}| \cdot K \cdot n \cdot n'\right).
%     \label{eq_complexity_readout}
% \end{equation}

\subsubsection{Overall Complexity and Analysis}
Combining the three stages, the total per-sample inference complexity of GSID is
\begin{equation}
    \mathcal{O}\!\left(|\mathcal{V}| \cdot |\mathcal{M}| \cdot n 
        \;+\; L \cdot |\mathcal{E}| \cdot n^2
        % \;+\; |\mathcal{V}| K n n'
        \right).
    \label{eq_complexity_total}
\end{equation}
In practical network topologies, the dominant part of the complexity is the message-passing stage, i.e., $\mathcal{O}(L |\mathcal{E}| n^2)$, which is linear to the number of edges and quadratic in the hidden dimension $n$. 
% This is consistent with the standard GATv2 complexity~\cite{GATv2}, with the additional constant factor $|\mathcal{T}|$ absorbed by the per-edge-type weight matrices $\{\mathbf{W}_\tau\}$. Since real-world routing topologies are sparse (i.e., $|\mathcal{E}| = \mathcal{O}(|\mathcal{V}|)$ for typical BGP/OSPF deployments), GSID scales quasi-linearly with the number of routers $|\mathcal{V}|$, making it well-suited for deployment in large-scale operational networks.
{The computational overhead of IDA is identical to that of the standard GATv2~\cite{GATv2}, as both apply a linear transform to the concatenated endpoint embeddings followed by a dot product with a learnable vector; the additional constant factor $|\mathcal{T}|$ introduced by the edge-type-specific weight matrices $\{\mathbf{W}_\tau\}$ does not affect the asymptotic complexity.
In terms of memory overhead, each message-passing layer stores the weight matrices $\{\mathbf{W}_\tau\}_{\tau\in\mathcal{T}}$ and attention vectors $\{\mathbf{a}_\tau\}_{\tau\in\mathcal{T}}$, incurring $\mathcal{O}(|\mathcal{T}| \cdot n^2)$ parameter memory per layer, which does not rely on the network scale $|\mathcal{V}|$. 
Since real-world routing topologies are sparse (i.e., $|\mathcal{E}| = \mathcal{O}(|\mathcal{V}|)$ for typical BGP/OSPF deployments) according to~\eqref{eq_complexity_total}, the inference cost of GSID scales quasi-linearly with the number of network nodes $|\mathcal{V}|$, making it practical for deployment in large-scale operational networks.}

% % ====================================================================
% \begin{figure*}[t]
%     \centering
%     \includegraphics[width=0.99\linewidth]{figs/sims/fig_overall_config.pdf}
%     \caption{Training results (Overall Performance).}
%     \label{fig_training_overall}
% \end{figure*}
% % ====================================================================

% ====================================================================
\begin{figure*}[b]
    \centering
    \includegraphics[width=\textwidth]{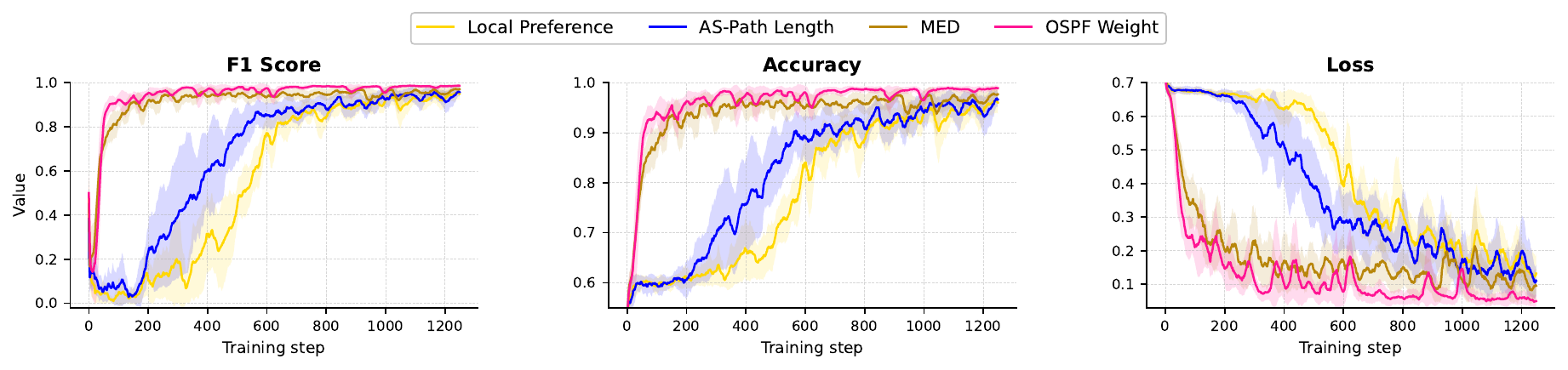}
    \caption{Training convergence of GSID for the four configurable parameters. The four features converge sequentially, reflecting their intrinsic differences in diagnostic difficulty.}
    \label{fig_training_40_error_rate_individual_features}
\end{figure*}
% ====================================================================

% ====================================================================
\begin{figure*}[t]
    \centering
    \includegraphics[width=\textwidth]{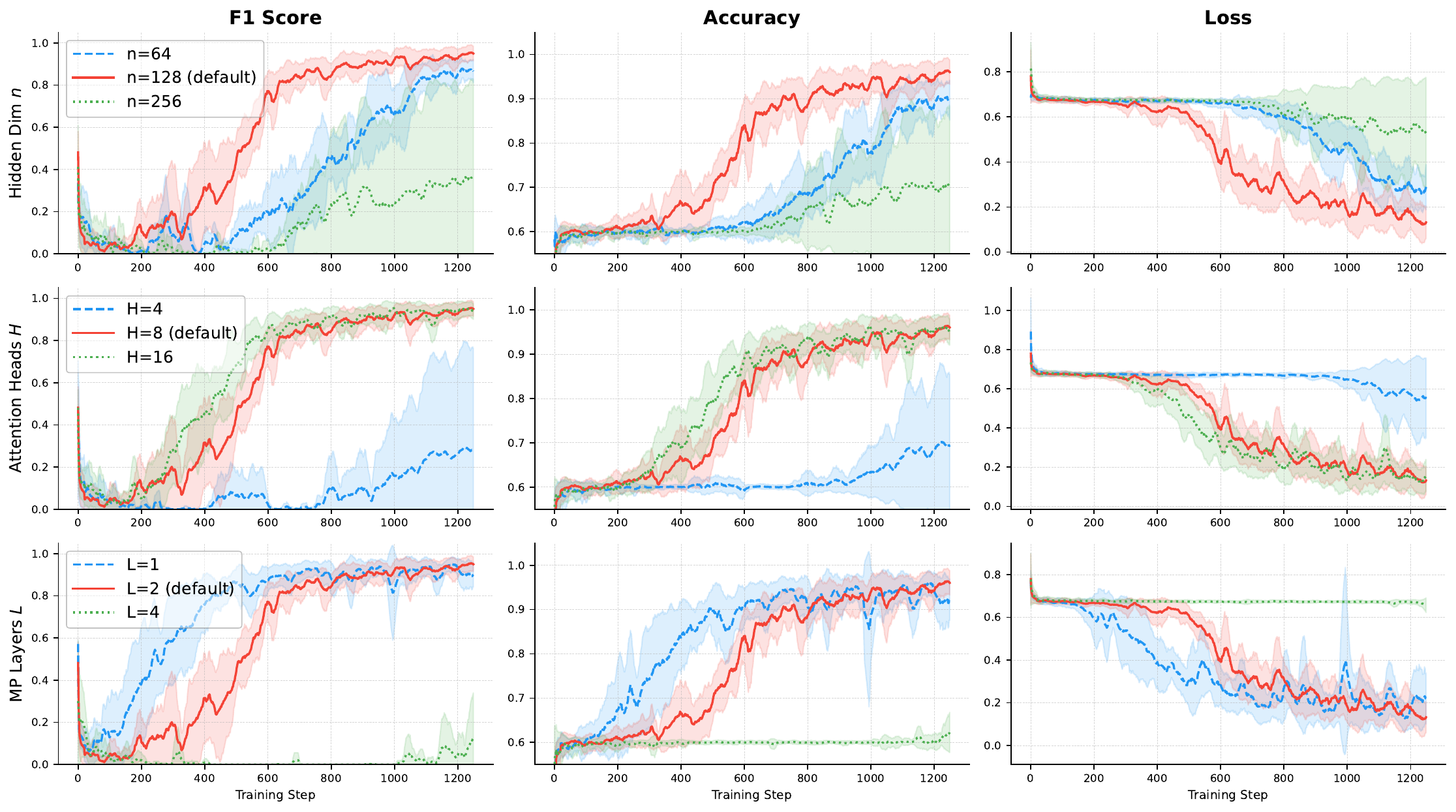}
    \caption{{Sensitivity analysis of GSID on local preference across three hyperparameters: the number of hidden dimensions, $n$ (top), the number of attention heads $H$ (middle), and the number of message-passing layers $L$ (bottom). The default configuration ($n=128$, $H=8$, $L=2$) consistently achieves high F1 score, accuracy, and training loss across all three metrics, confirming that it is well-suited for detecting protocol configuration anomalies in the most diagnostically challenging parameter.}}
    \label{fig_sensitivity_local_preference}
\end{figure*}
% ====================================================================

% ====================================================================
\begin{figure*}[t]
    \centering
    \includegraphics[width=\textwidth]{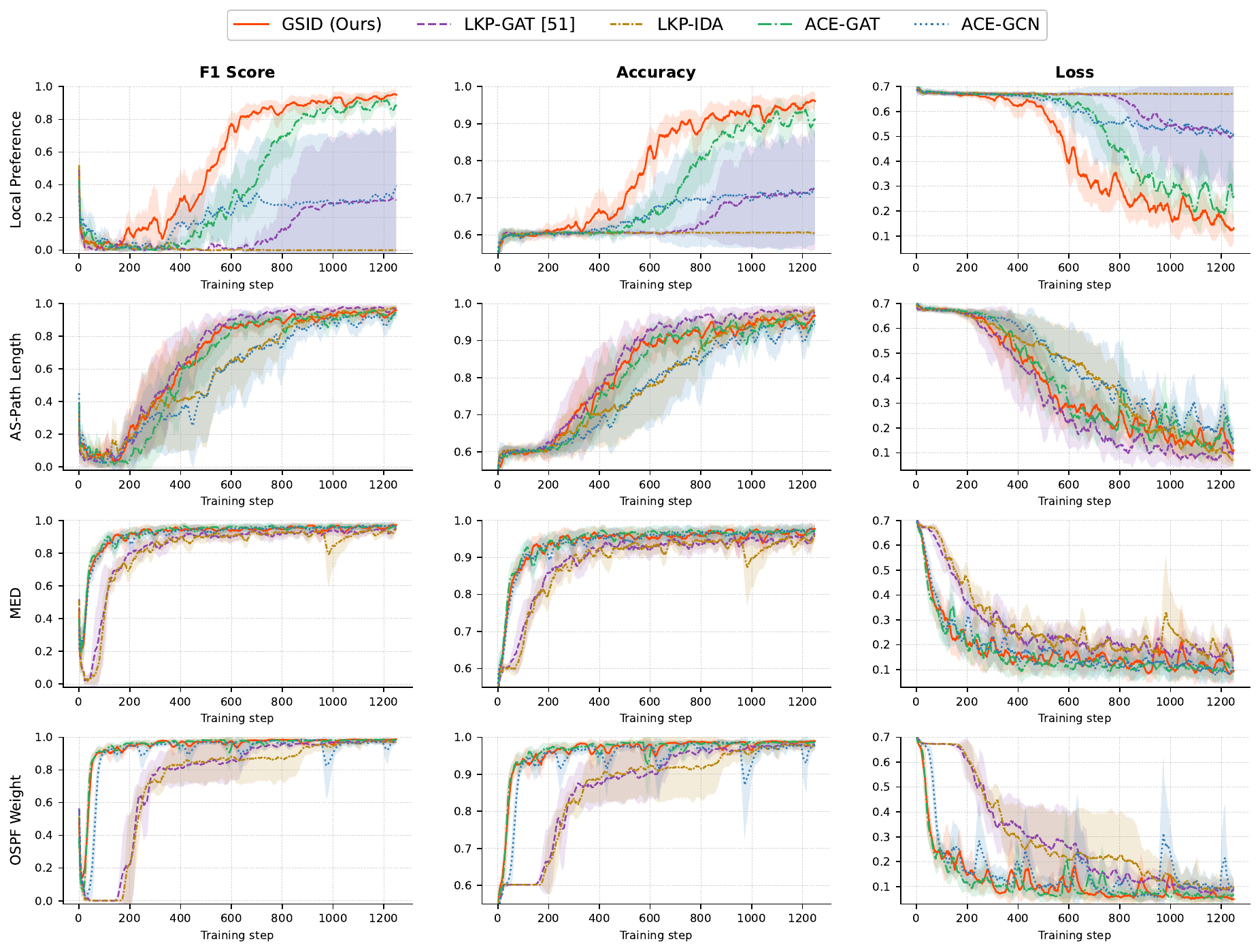}
    \caption{Training curves across all protocol configuration parameters. The proposed GSID consistently outperforms the baseline algorithms, with the performance gap most pronounced for local preference. In addition, GSID demonstrates higher robustness, evident from the narrower confidence intervals across runs, especially when compared with the LKP-GAT baseline and ACE-GCN benchmark.}
    \label{fig_training_40_error_rate_compare_algs}
\end{figure*}
% ====================================================================

\section{Performance Evaluation and Analysis}\label{section_performance_evaluation}
In this section, we gauge the convergence of the proposed GSID algorithm, followed by a comparative ablation study against state-of-the-art baselines in detection accuracy, F1 score, and training loss. We also evaluate the adaptation capability of GSID under unseen conditions, i.e., different datasets that differ in topology structure, network scale, and anomaly rate, by {OOD generalization}.

\subsection{Experimental Setup}\label{section_exp_setup}
To construct the datasets for training and testing, we employ the BGP and OSPF protocol simulator developed in~\cite{NeurIPS_BGP}. This simulator generates bipartite graph representations for network samples under specified parameters and protocols. Specifically, each network sample is generated with a randomly drawn number of 16-23 internal routers, 3 gateway nodes, and 4-7 destination networks. The number of routing policy constraints is also randomized: 8-12 forwarding (\texttt{fwd}) rules, 4-7 reachability (\texttt{reachable}) rules, and 10-30 traffic isolation (\texttt{trafficIso}) rules. We evaluate GSID on the transformed bipartite graphs from the generated network samples. {For the real-world topology experiments, the physical topology structures are derived from operational networks via the Internet Topology Zoo~\cite{TopologyZoo}, while the protocol configuration parameters are synthetically generated due to limited access to practical operational network configuration data in practice.}

To simulate protocol configuration anomalies, we inject anomalies into the four numerical configuration parameters: 
$\theta_v^{\mathrm{LP}}$, $\theta_{v,u}^{\mathrm{ASL}}$, $\theta_{v,u}^{\mathrm{MED}}$, and $\phi_{v,u}$. 
It was reported in~\cite{SIGCOMM_Understand_BGP_Misconfig} that up to 75\% of advertised BGP prefixes can be misconfigured.
By default, for each parameter, we randomly select $40\%$ of its eligible fact nodes and replace them with a different value drawn uniformly from the same protocol-defined range, e.g., $[\theta^{\mathrm{LP}}_{\min}, \theta^{\mathrm{LP}}_{\max}]$.
The anomaly label is set to $y_{v,c}=1$ if the $c$-th parameter of the $v$-th node is perturbed, i.e., injected with a different value drawn uniformly randomly from the protocol-defined valid range; and $y_{v,c}=0$, otherwise; see \eqref{eq_anomaly_label}. With the GSID model trained on a $40\%$ anomaly injection rate, the model is {evaluated under OOD generalization settings} under unseen network scenarios with a wide spectrum of anomaly rates ranging from 1\% to 80\%.

All models are trained with the Adam optimizer (weight decay $10^{-5}$) for $400$ epochs, with a batch size of $4$. The learning rate is initialized at $5 \times 10^{-4}$ and linearly decayed to $1 \times 10^{-4}$ over the first $10$ epochs, then held constant thereafter. All models use the same training dataset comprising 4,096 networks (and bipartite graphs) share the same architecture hyperparameters: hidden dimension $128$, $8$ attention heads, $2$ message-passing encoder layers followed by $3$ iterative decoder layers, and dropout rate $0.2$. To address different convergence difficulty levels across the four monitored protocol features, we adopt dynamic weight averaging with temperature $T = 2.0$ to balance the per-feature cross-entropy losses during training. All experiments are repeated three times. The reported curves show the means with the shaded areas representing the standard deviation across runs. Please refer to the GitHub link \href{https://github.com/hxheart/GSID}{https://github.com/hxheart/GSID}. 

% {{For some reason, we also use in the implementation [ref of my invited paper].} In communication networks, different protocols exhibit distinct alarm propagation patterns. For example, OSPF weight anomalies propagate through link-state updates along $\mathcal{E}_{\text{OSPF}}$ edges, affecting shortest-path computations. However, BGP local preference errors spread via $\mathcal{E}_{\text{iBGP}}$ sessions, altering route selection decisions.}
 
\subsection{Convergence {and Sensitivity} of GSID}\label{section_training_convergence}
\subsubsection{{Convergence Performance}}
Fig.~\ref{fig_training_40_error_rate_individual_features} shows that GSID converges steadily across all four configurable parameters of F1 score, accuracy, and training loss. We can also observe that the four protocol parameters converge sequentially: OSPF weight $\phi_{i,j}$ converges fastest, followed by MED $\theta^{\text{MED}}_{i,j}$, AS path length $\theta^{\text{ASL}}_{i,j}$, and local preference $\theta^{\text{LP}}_v$, which exhibits the slowest convergence. Interestingly, this ordering aligns with the scope of influence of each parameter, from a single link ($\phi_{i,j}$) to inter-AS route selection ($\theta^{\text{MED}}_{i,j}$, $\theta^{\text{ASL}}_{i,j}$) to AS-wide exit-point selection ($\theta^{\text{LP}}_v$), which is consistent with the BGP decision priority established in Section~\ref{section_protocol_configuration_model}. This suggests that higher-priority parameters produce more widespread anomaly symptoms that are harder to attribute, and hence demand more training steps to converge.

% ########################################################################################################

\begin{table}[t]
\centering
\caption{{Hyperparameter Ranges for Sensitivity Analysis}}
\label{tab:sensitivity}
\begin{tabular}{lrr}
\toprule
\textbf{Hyperparameter} & \textbf{Default} & \textbf{Range Tested} \\
\midrule
Hidden dimension $n$       & 128 & \{64, \textbf{128}, 256\} \\
Attention heads $H$        & 8   & \{4,  \textbf{8},   16\}  \\
Message-passing layers $L$ & 2   & \{1,  \textbf{2},   4\}   \\
\bottomrule
\end{tabular}
\end{table}

\subsubsection{Sensitivity Analysis}
To assess the robustness of GSID, we conduct a sensitivity analysis over three key hyperparameters: the number of hidden dimensions $n$, the number of attention heads $H$, and the number of message-passing layers $L$. These parameters collectively govern the model's representational capacity, its ability to capture multi-scale relational patterns, and the depth of neighborhood aggregation. The tested ranges for each hyperparameter are summarized in Table~\ref{tab:sensitivity}. 
We focus on local preference $\theta^{\mathrm{LP}}$, since it is the most diagnostically challenging parameter, evidenced by its slowest convergence among all four protocol features in Fig.~\ref{fig_training_40_error_rate_individual_features}. The corresponding AS-wide scope of influence indicates that even small hyperparameter variations of $\theta^{\mathrm{LP}}$ are likely to manifest as observable performance differences, making it the most informative setting for validating hyperparameter choices. The remaining three parameters converge more readily and exhibit lower sensitivity to hyperparameter variations.

Fig.~\ref{fig_sensitivity_local_preference} shows that the default configuration ($n=128$, $H=8$, $L=2$) achieves consistently favorable performance across all three metrics. Reducing the number of hidden dimensions to $n=64$ leads to noticeably degraded F1 scores and slower convergence; a moderate embedding capacity is necessary to preserve the fine-grained numerical structure required for detecting local preference anomalies. Increasing it to $n=256$ yields marginal improvement at the cost of higher computational overhead, making $n=128$ the preferred choice. A similar trend is observed for the number of attention heads: $H=4$ results in slower convergence and higher variance across runs, while $H=16$ yields a slight improvement in convergence speed without a meaningful gain in final performance, making $H=8$ the preferred choice, balancing efficiency and performance. With respect to the number of message-passing layers, $L=1$ limits the receptive field of each node, hampering the model's ability to aggregate the multi-hop context that local preference anomalies require, whereas $L=4$ shows signs of over-smoothing that slightly degrades final performance. These observations confirm that the default hyperparameter configuration strikes a balance between model expressiveness and generalization for protocol anomaly detection.

% \end{color}

% ########################################################################################################
\subsection{Benchmarks}
We compare GSID with the state of the art, and conduct ablation studies by individually replacing its components.

% \begin{itemize}
%     \item \textit{LKP-GAT} (state-of-the-art baseline adapted from~\cite{NeurIPS_BGP}): This baseline uses the lookup encoder for the four configurable parameters with GAT's static attention mechanism for neighborhood aggregation, and is extended from \cite{NeurIPS_BGP}. 
    % a most recently and widely used method for BGP and OSPF configuration synthesis. 
% it is adaptable to our anomaly detection task by replacing its configuration synthesis head with a node-level binary classifier. However,
\subsubsection{State-of-the-art Baseline}
The baseline, \textit{LKP-GAT}, is adapted from~\cite{NeurIPS_BGP}.
% which addresses a configuration synthesis problem using a lookup-based encoder with GAT's static attention. 
Notably, the study in \cite{NeurIPS_BGP} addresses a different configuration synthesis problem with the configurable BGP and OSPF parameters.
% , making it a representative baseline for our anomaly detection task. 
Unlike GSID, its lookup encoder maps each configuration parameter to a learnable embedding via a one-hot projection; static attention scores are computed using nonlinearity to each node's individual feature before attention. Thus, the capability of LKP-GAT to capture the structural features of connected nodes and edges is limited.

% reason the optimal configuration parameters

\subsubsection{Ablation Variants}
Three reduced variants of GSID are considered to assess the contributions of its key components.
\begin{itemize}
    \item \textit{LKP-IDA}: retains the IDA mechanism, and replaces the ACE numerical encoder with a standard lookup encoder, isolating the contribution of the ACE;
    \item \textit{ACE-GAT}: retains the ACE numerical encoder, and replaces the IDA with GAT's static attention mechanism~\cite{GAT_ICLR_2018}, isolating the gain of IDA mechanism;
    \item \textit{ACE-GCN}: retains the ACE numerical encoder, and replaces the attention mechanism with GCN's uniform neighborhood aggregation~\cite{GCN_classical}, where all neighbors contribute equally to the aggregated message,
    % without explicitly distinguishing the spatial variation, 
    serving as a non-attention ablation.
\end{itemize}

% ====================================================================
% \begin{figure*}[t]
%     \centering
%     \includegraphics[width=\textwidth]{figs/sims/TNSE/fig_GATv2_NUM_per_feature_20_error_rate.pdf}
%     \caption{Training convergence of GSID for each of the four monitored protocol configuration parameters, under a 20\% anomaly injection rate. The four features converge sequentially, consistent with the 40\% injection rate results in Fig.~\ref{fig_training_40_error_rate_individual_features}.}
%     \label{fig_training_20_error_rate_individual_features}
% \end{figure*}
% ====================================================================
 
% ====================================================================
\begin{figure*}[t]
    \centering
    \includegraphics[width=\textwidth]{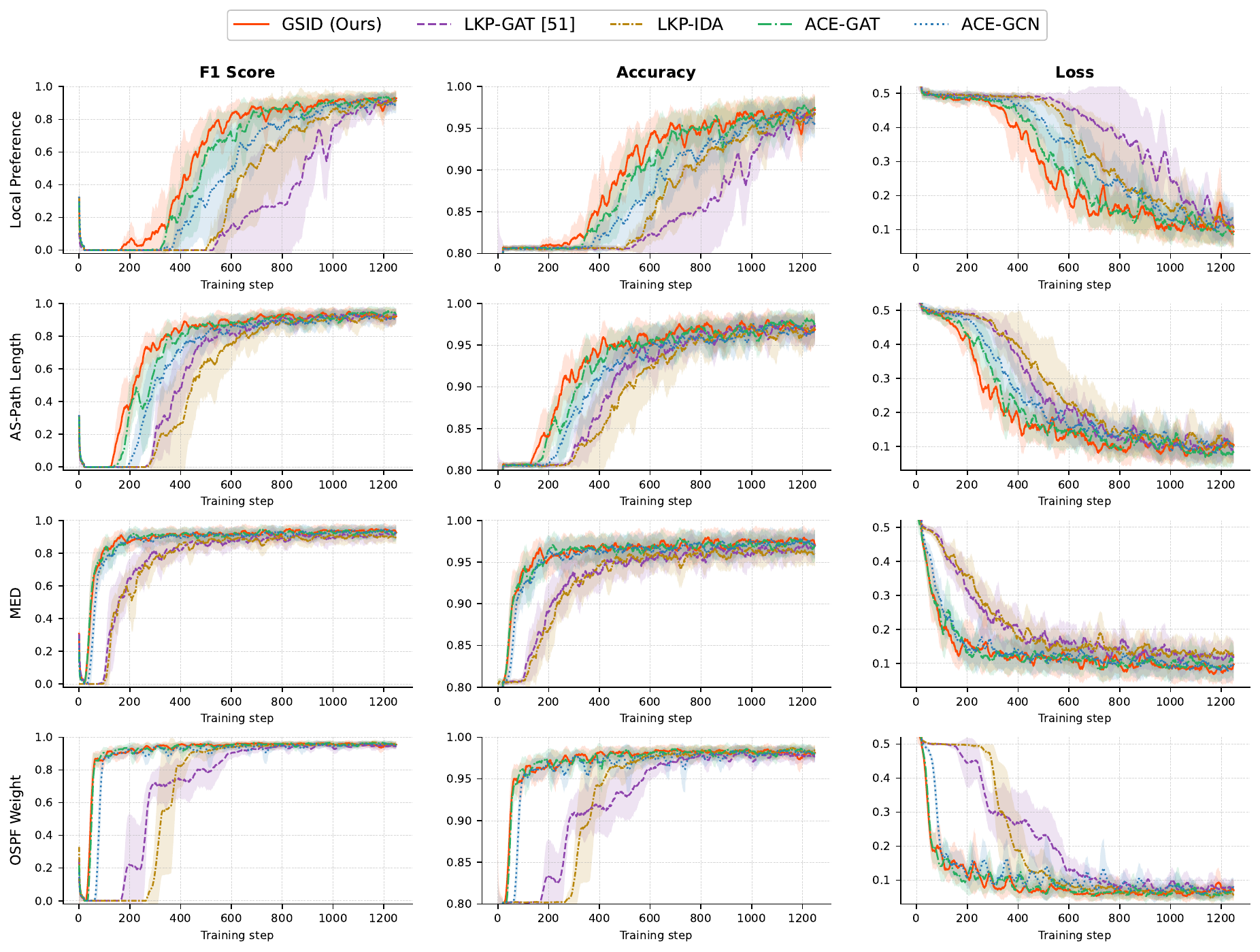}
    \caption{Training curves across all protocol configuration parameters at a 20\% anomaly injection rate. GSID consistently achieves the highest F1 score and accuracy with the fastest convergence across all four features and all three metrics, although all five algorithms can converge in this scenario.}
    \label{fig_training_20_error_rate_compare_algs}
\end{figure*}
% ====================================================================

\subsection{Training Comparison with Baselines and Ablation Study}

% 

% \subsubsection{Training Results}\label{section_training_ablation}
Fig.~\ref{fig_training_40_error_rate_compare_algs} compares GSID with the LKP-GAT baseline and three ablation variants across all four protocol features.

\subsubsection{Comparison with Baseline}
Comparing GSID against LKP-GAT baseline~\cite{NeurIPS_BGP}, GSID achieves higher F1 scores across all four protocol features. For local preference $\theta^\mathrm{LP}$, GSID reaches a final F1 of $93.2\%$ versus LKP-GAT's $30.4\%$, improving F1 score by more than 3 times, and achieves an accuracy of $94.8\%$ versus LKP-GAT's $71.5\%$, improving accuracy by 23.2\%. GSID also exhibits narrower confidence bands and faster convergence than the LKP-GAT baseline, evident from the narrower confidence intervals across runs. For instance, on OSPF weight $\theta^\mathrm{OSPF}$, GSID reaches $90\%$ of LKP-GAT's converged F1 in only $62$ steps versus $637$, reducing the training steps by approximately $90.3\%$. GSID provides both higher reliability and training efficiency than the baseline, which are critical for detecting protocol anomalies whose symptoms exhibit spatial variation and locality.

\subsubsection{Ablation Studies}
The following three observations isolate the contributions of the ACE and the IDA in GSID.

\textit{First, the proposed ACE provides consistent gains over the lookup-based baseline.}
Comparing GSID with LKP-IDA, which isolates the contribution of the CA node feature encoder proposed in Section~\ref{section_design_encoder}. The proposed GSID consistently outperforms LKP-IDA across all four protocol features. Notably, LKP-IDA fails to detect anomalies in local preference, resulting in a near-zero F1 score throughout training. This demonstrates the effectiveness of our ACE encoding, and validates that the proposed configuration-aware encoding strategy is critical for preserving the fine-grained numerical structure that protocol anomaly detection requires.

\textit{Second, attention mechanisms are essential for protocol anomaly detection.} 
Comparing GSID against ACE-GCN isolates the contribution of any attention mechanism. ACE-GCN fails to converge for local preference. This is consistent with the diagnostic difficulty discussed in Fig.~\ref{fig_training_40_error_rate_individual_features}, where local preference exhibits the slowest convergence across the network topology, confirming that attention-based mechanisms are essential for protocol anomaly detection tasks. 

\textit{Third, the proposed IDA provides a decisive advantage over static attention.} 
Comparing GSID against ACE-GAT isolates the contribution of the IDA mechanism proposed in Section~\ref{section_design_GATv2}. ACE-GAT converges more slowly and achieves lower final F1 scores than GSID, particularly for local preference. This confirms the effectiveness of the joint-state dependency discussed in Section~\ref{section_design_GATv2}, where the diagnostic relevance of a neighboring node depends on the combined state of both endpoints rather than each node independently.

% ==== table for topologies 

% {

\subsubsection{Adaptivity to Different Anomaly Rates}
We give training results for the 20\% anomaly injection rate in Fig.~\ref{fig_training_20_error_rate_compare_algs}.
% Fig.~\ref{fig_training_20_error_rate_individual_features} and~\ref{fig_training_20_error_rate_compare_algs}.

% \paragraph{Convergence of GSID}
% Fig.~\ref{fig_training_20_error_rate_individual_features} shows the training convergence of GSID under a $20\%$ anomaly injection rate. The sequential convergence pattern is preserved as in Fig.~\ref{fig_training_40_error_rate_individual_features}, with OSPF link weight converging fastest, followed by AS-path length and MED at an intermediate pace, while local preference exhibits the slowest and most volatile trajectory. This confirms that the convergence sequence reflects the intrinsic diagnostic difficulty of each protocol feature rather than the statistical prevalence of anomalous samples. 

% \paragraph{Comparison with Baseline and Ablation}
Fig.~\ref{fig_training_20_error_rate_compare_algs} shows that the relative ordering of all algorithms is preserved across all four protocol features, with GSID consistently achieving the highest F1 score and accuracy. Unlike the training results with $40\%$ anomaly rates, under this lower anomaly injection rate, GSID and the four benchmarks can all converge at the end of training, indicating that a lower anomaly rate reduces the overall anomaly detection difficulty. Nevertheless, the performance gap between GSID and the benchmarks is preserved, confirming that the superiority of GSID stems from our algorithmic design.

% ================================================================
\begin{table}[t]
\caption{Key Parameters of Different Datasets}
\label{tab_dataset_configurations}
\centering
\setlength{\tabcolsep}{3pt}
\renewcommand{\arraystretch}{1.3}
{
% \footnotesize
\begin{tabular}{l|rrr}
\hline
\hline
\multicolumn{1}{l|}{\diagbox[width=10.6em]{\textbf{Parameters}}{\textbf{Networks}}} & \textbf{Baseline} & \textbf{~~Larger-Scale} & \textbf{Real-World}
% \vspace{2pt}
\\
\hline
Topology Type                   & Synthetic         & Synthetic             & ~~~Topology Zoo 
% \vspace{2pt}
\\
\hline
Router Numbers                  & 16--23            & 24--31                & By scenarios \\
Dest. Network Numbers           & 4--7              & 10--15                & 4--7 \\
Gateway Node Numbers            & 3                 & 7--9                  & 3 \\
\texttt{fwd} Querie Numbers     & 8--12             & 25--35                & 8--12 \\
\texttt{reach} Querie Numbers   & 4--7              & 15--20                & 4--7 \\
\texttt{iso} Querie Numbers     & 10--30            & 10--30                & 10--30 
\vspace{2pt}
\\
\hline
Train Sample Numbers            & 4096              & --                    & -- \\
Test Sample Numbers             & 100               & 100                   & 100 \\
\hline
\hline
\end{tabular}
}
\end{table}
% ====================================================================

% ==== bar figures ===================================================
\begin{figure}[t]
    \centering
    \includegraphics[width=0.49\textwidth]{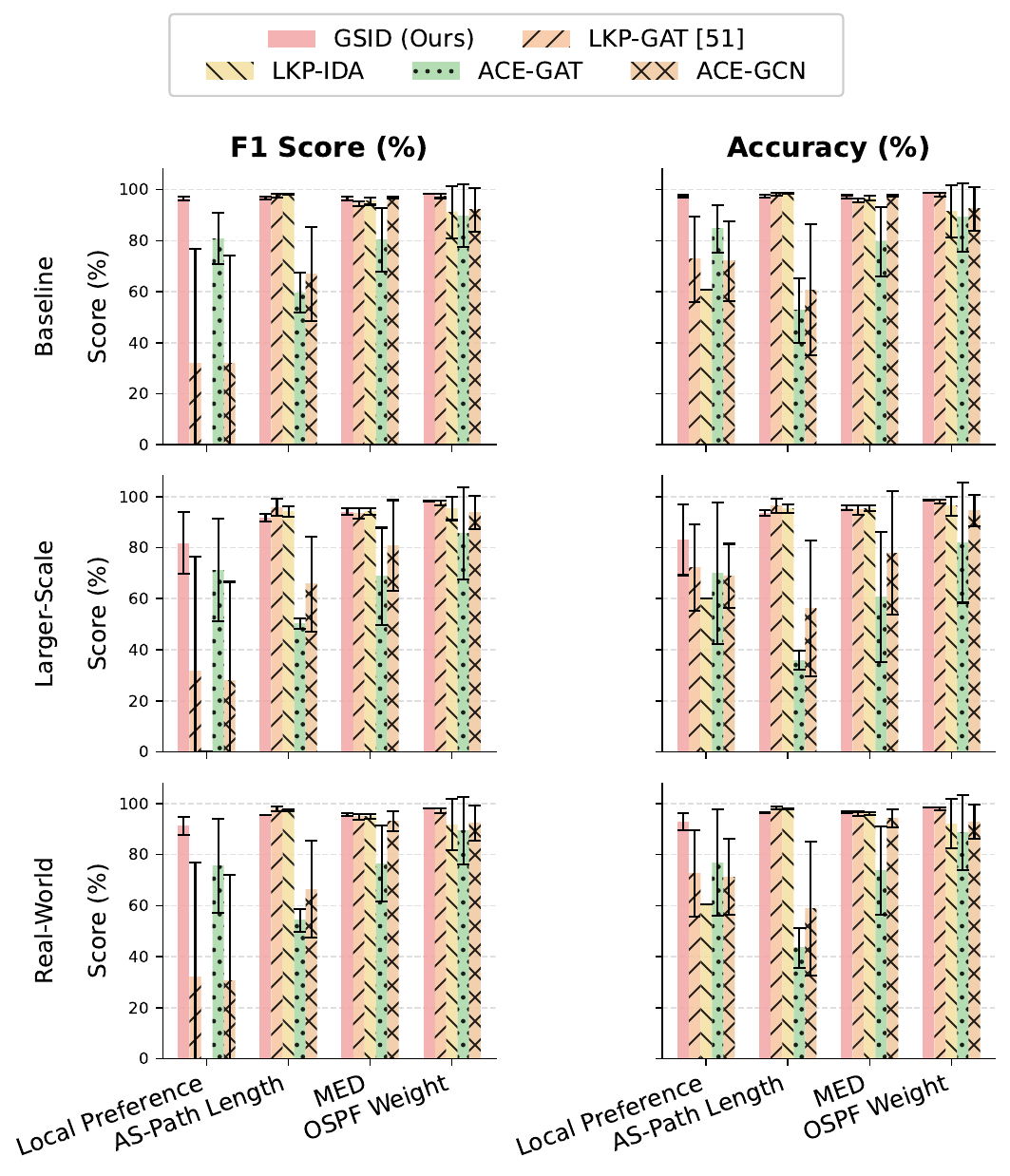}
    \caption{Scalability across different topologies. Validation across topology variations. From left to right are the baseline, larger-scale, and real-world datasets that are summarized in Table~\ref{tab_dataset_configurations}, with 100 validation samples in each.}
    \label{fig_test_different_network_scale_test_40_by_train_40}
\end{figure}
% ====================================================================

% % ====================================================================
% \begin{figure}[t]
%     \centering
%     \includegraphics[width=0.49\textwidth]{figs/sims/TNSE/fig_test_different_network_scale_test_20_by_train_40.pdf}
%     \caption{Testing results of anomaly injection rate 20\% by using the trained model with anomaly injection rate 40\%.}
%     \label{fig_test_different_network_scale_test_20_by_train_40}
% \end{figure}
% % ====================================================================

% ---------- Figure: Performance vs. Test Error Rate ----------
\begin{figure*}[!t]
    \centering
    \begin{subfigure}[b]{0.49\textwidth}
        \includegraphics[width=\linewidth]{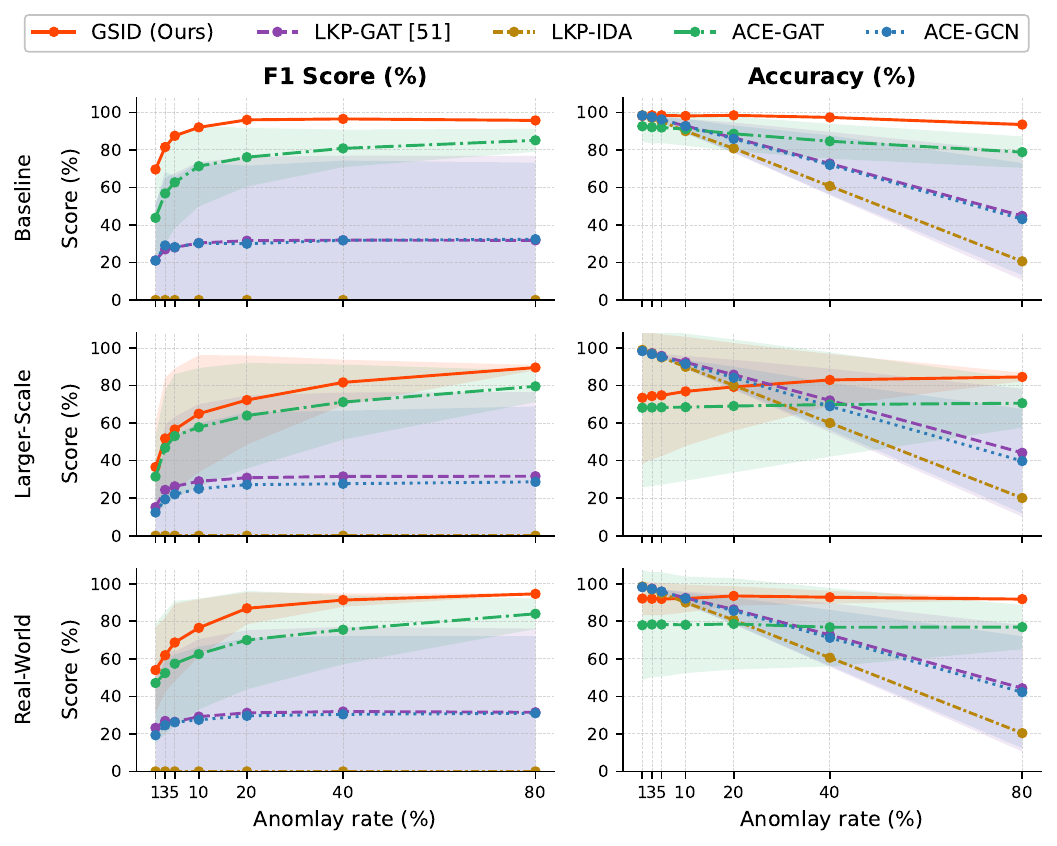}
        \caption{Local preference}
        \label{fig:trend_lp}
    \end{subfigure}
    \hfill
    \begin{subfigure}[b]{0.49\textwidth}
        \includegraphics[width=\linewidth]{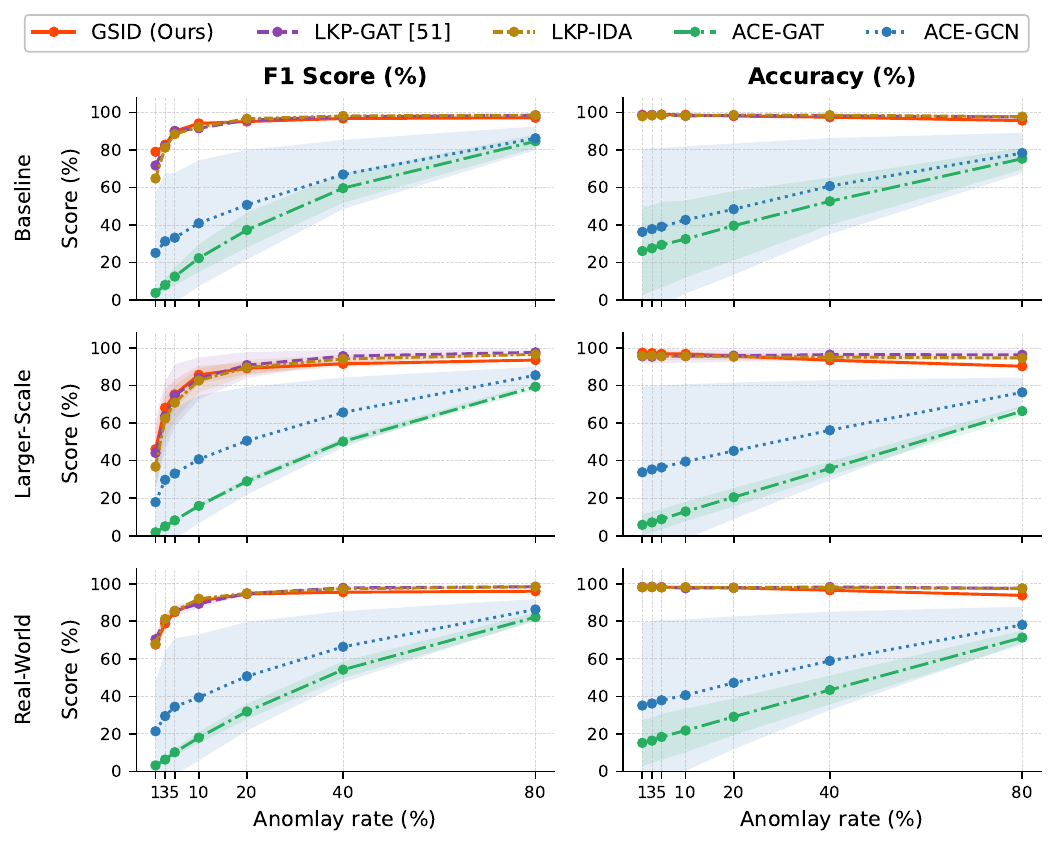}
        \caption{AS-path length}
        \label{fig:trend_aspath}
    \end{subfigure}
    
    \vspace{0.5em}
    
    \begin{subfigure}[b]{0.49\textwidth}
        \includegraphics[width=\linewidth]{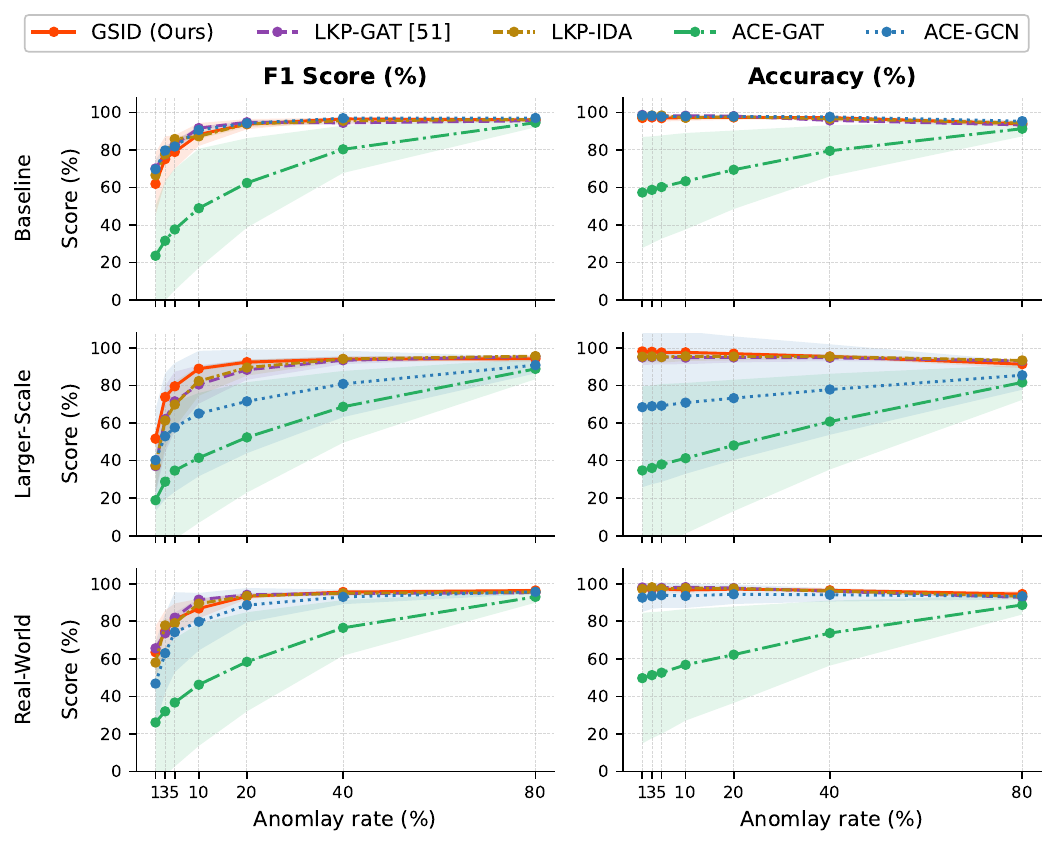}
        \caption{MED}
        \label{fig:trend_med}
    \end{subfigure}
    \hfill
    \begin{subfigure}[b]{0.49\textwidth}
        \includegraphics[width=\linewidth]{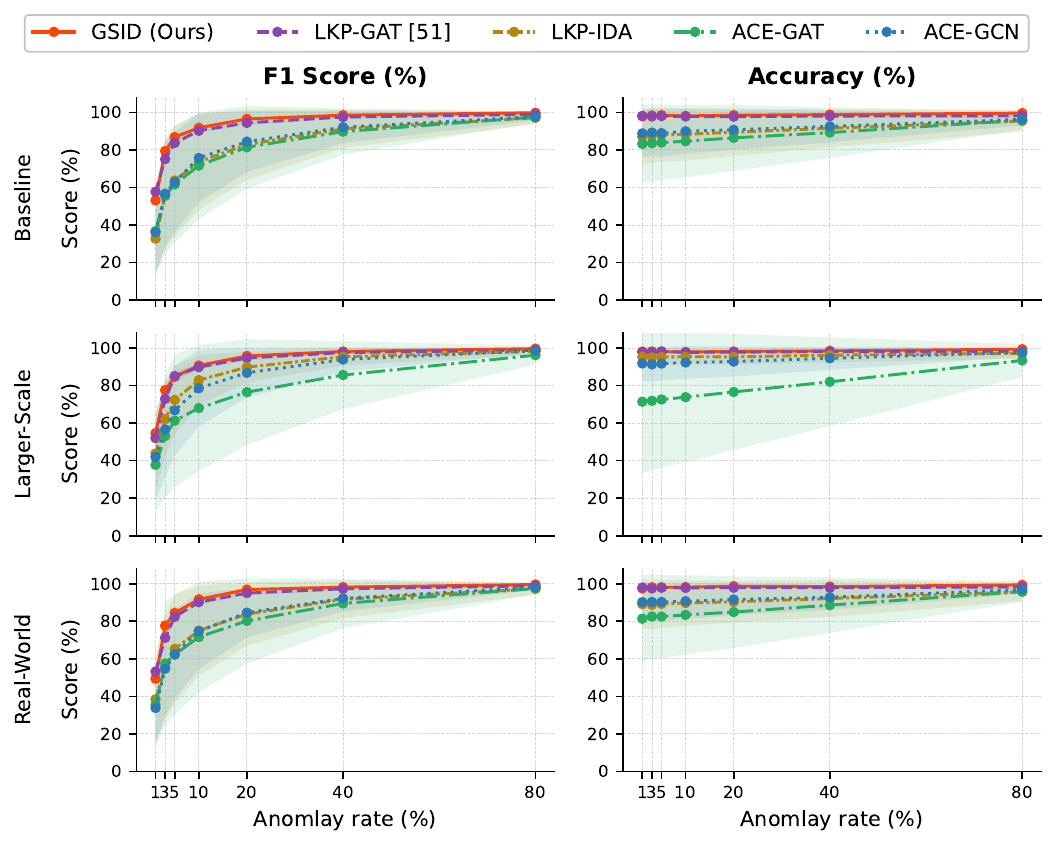}
        \caption{OSPF weight}
        \label{fig:trend_ospf}
    \end{subfigure}
    \caption{Performance of all five algorithms as a function of anomaly rate, across three network scales and two metrics. For local preference, some benchmarks achieve unacceptably low F1 scores on large-scale and real-world datasets under low injection rates, despite their high accuracy inflated by class imbalance.
    % Each panel corresponds to one target feature
    % {explain why some algorithms can have extremely high accuracy, but sacrifice F1 score, caused by the data imbalance}.
    }
    \label{fig_trend_vs_error_rate}
\end{figure*}

\subsection{Testing Results}
We evaluate the adaptive capability of GSID under three out-of-distribution conditions with increasing difficulty: unseen network topologies at the training scale, unseen larger-scale network topologies, and unseen real-world topologies from the Internet Topology Zoo~\cite{TopologyZoo}. In addition, we evaluate robustness to unseen anomaly injection rates. All models are trained only on the baseline dataset with the default 40\% anomaly injection rate, and evaluated {under OOD generalization} across the anomaly rate of 1\% -- 80\%. The key parameters of the three datasets are summarized in Table~\ref{tab_dataset_configurations}.

\subsubsection{Adaptivity to Unseen Topologies, Scales, and Real-World Networks}
Fig.~\ref{fig_test_different_network_scale_test_40_by_train_40} shows F1 score and accuracy across three dataset regimes under the same anomaly injection rate as training. GSID maintains strong performance across all three settings compared with the four benchmark algorithms.

On the baseline testing dataset, LKP-IDA completely fails to detect anomalies in local preference, since it achieves a zero F1 score throughout, highlighting the critical role of the ACE in preserving the fine-grained deviation signals. The ACE-GAT benchmark, despite retaining the ACE, achieves lower F1 scores than GSID, particularly for local preference $\theta^\mathrm{LP}$, where GSID attains $95.7\pm0.9\%$ F1 with minimal run-to-run variance, confirming that static attention is insufficient to capture the joint-state dependencies among neighboring nodes. The ACE-GCN baseline similarly struggles with local preference, confirming that both adaptive encoding and attention-based aggregation are indispensable for detecting the configuration anomalies.

On the larger-scale dataset, GSID demonstrates far more reliability compared with the LKP-GAT baseline, whose F1 on local preference is approximately $31.6\pm44.7\%$. Among the baselines, LKP-IDA continues to fail on local preference, and ACE-GAT and ACE-GCN exhibit more pronounced degradation than GSID, indicating that static attention and uniform aggregation are increasingly inadequate as the network scale grows and diagnostic signals must propagate over longer paths.

On the real-world topology dataset from the Internet Topology Zoo~\cite{TopologyZoo}, GSID outperforms all baselines across four protocol features. On local preference $\theta^\mathrm{LP}$, GSID attains an F1 of $85.6\pm6.3\%$, whereas LKP-GAT collapses to $31.9\pm45.1\%$, demonstrating the reliability of our GSID. Real-world network topologies exhibit irregular degree distributions, asymmetric peering structures, and diverse AS boundary configurations that are absent from the synthetic training graphs. The observation that GSID generalizes well to these topologies validates that the structural priors encoded in its message passing capture genuine protocol patterns. ACE-GAT and ACE-GCN show larger performance drops, particularly for local preference and MED, whose diagnostic signals span longer propagation paths that are more topology-dependent under real-world irregular structures. LKP-IDA again achieves near-zero F1 on local preference, confirming that numerical encoding is a prerequisite for reliable anomaly detection, irrespective of the underlying topology.

\subsubsection{Adaptivity to Unseen Anomaly Rate}
Fig.~\ref{fig_trend_vs_error_rate} presents the {OOD generalization} F1 score and accuracy of all algorithms as a function of anomaly injection rate, ranging from $1\%$ to $80\%$, across the three network scales of baseline, large-scale, and real-world. GSID consistently achieves the highest or near-highest performance across all four protocol features and all datasets. As the injection rate decreases, F1 scores of all algorithms decline due to increased data class imbalance, but GSID exhibits the most graceful degradation. 

This imbalance effect further manifests as a spurious accuracy advantage observed at low injection rates. Although LKP-GAT, LKP-IDA, and ACE-GCN attain comparably high accuracy on local preference $\theta^{\mathrm{LP}}$ at low injection rates, particularly on the large-scale and real-world datasets, their unacceptably low F1 scores reveal failures in detecting true anomalies. This is because when anomalous samples are extremely sparse, these models trivially exploit the class imbalance by predicting the normal class. {To mitigate this effect, GSID adopts dynamic weight averaging with temperature $T = 2.0$ to balance the cross-entropy losses across all monitored configuration parameters during training, which contributes to its more graceful performance degradation compared with the baselines.}

\section{Conclusion and Future Works}\label{section_conclusion}
In this paper, we modeled the communication network as a bipartite graph, and formulated the problem of detecting protocol configuration anomalies in sovereign network functions (particularly, BGP and OSPF) as the detection of structural and semantic inconsistencies of connected nodes and edges in the graph. We proposed GSID with two meticulously designed modules, an ACE and an IDA, to encode diverse protocol parameter features and excavate subtle inconsistencies in the bipartite graph, respectively. Training results showed that GSID outperforms the baselines by 3 times in F1 score and by 23.2\% in accuracy. {OOD generalization} testing on unseen network scales and anomaly injection rates, and real-world topologies validated the superior adaptability of GSID.
As future work, we plan to extend the network and anomaly models to more complex operational error patterns and explicit specification-configuration mappings, and design label-free algorithms like self-supervised learning. {Furthermore, the protocol configuration parameters in the current evaluation remain synthetically generated, as real-world network configuration data are subject to practical constraints on availability, privacy, and operational sensitivity. The incorporation of real-world configuration traces, including BGP routing data collected from live networks and validation against real-world anomaly events, will be an important direction for future work.}

% ====================================================================
% Appendix
% ====================================================================
\appendices

\renewcommand{\theequation}{A.\arabic{equation}}
\setcounter{equation}{0}
\renewcommand{\thefigure}{A.\arabic{figure}}
\setcounter{figure}{0}
\renewcommand{\thetable}{A.\arabic{table}}
\setcounter{table}{0}

\section{Fact Node Predicate Structure}\label{appendix_predicate_structure}

Table~\ref{tab_fact_node_args} summarizes the argument structure of each fact node predicate and the corresponding edge types~$\tau$ used in the bipartite graph (cf.~Fig.~\ref{fig_bipartite_graph_representation}). Each fact node connects to its argument entities via directed edges whose type~$\tau \in \{0,1,2,3\}$ encodes the \emph{role} that the entity plays within the predicate, rather than merely its position in an argument list.  $\tau = 0$ identifies the \emph{primary subject} of the relationship (e.g.\ the source router in a forwarding rule), while $\tau = 1,2,3$ identify secondary arguments whose semantic role is predicate-specific.

The predicates fall into two groups. (i): \texttt{connected}, \texttt{iBGP}, \texttt{eBGP}, and \texttt{BGP\_route} are binary predicates ($\text{\# args} = 2$): each involves exactly one $\tau{=}0$ and one $\tau{=}1$ edge, capturing a symmetric or directed relationship between two entity nodes. (ii) \texttt{fwd} and \texttt{reachable} are ternary ($\text{\# args} = 3$), introducing a $\tau{=}2$ edge to a third entity (i.e., the next-hop router or exit router, respectively). Path information is preserved without flattening it into the binary edge structure. \texttt{trafficIso} is the sole quaternary predicate ($\text{\# args} = 4$), connecting two source routers and two destination networks through $\tau \in \{0,1,2,3\}$ edges to represent pairwise traffic-isolation requirements between two flows.

This typed-edge design keeps the graph structure uniform (a fact node is connected only to entity nodes) while allowing the GNN message-passing procedure to distinguish argument roles through the edge-type $\tau$; see
Sec.~\ref{section_graph_representation}.

% ============================================================
\begin{table}[t]
\centering
\caption{Argument structure of fact nodes and corresponding edge types~$\tau$
         (cf.\ Fig.~\ref{fig_bipartite_graph_representation}).
         R: router; N: network; ext.\ AS: external AS; —: not applicable.}
\label{tab_fact_node_args}
\renewcommand{\arraystretch}{1.3}
\begin{tabular}{lcccccc}
\toprule\toprule
\textbf{Fact node}   & \textbf{\# args} & $\bm{\tau}$\textbf{=0} & $\bm{\tau}$\textbf{=1} & $\bm{\tau}$\textbf{=2} & $\bm{\tau}$\textbf{=3} \\
\midrule
$\texttt{connected}(\cdot)$   & 2 & R       & R         & —  & — \\
$\texttt{iBGP}(\cdot)$        & 2 & R       & RR        & —  & — \\
$\texttt{eBGP}(\cdot)$        & 2 & R       & ext.\ AS  & —  & — \\
$\texttt{BGP\_route}(\cdot)$  & 2 & ext.\ AS & N        & —  & — \\
\hline
$\texttt{fwd}(\cdot)$         & 3 & R (src) & N (dst)   & R (next) & — \\
$\texttt{reachable}(\cdot)$   & 3 & R (src) & N (dst)   & R (exit) & — \\
$\texttt{trafficIso}(\cdot)$  & 4 & R$_1$   & R$_2$     & N$_1$    & N$_2$ \\
\bottomrule\bottomrule
\end{tabular}
\end{table}
% ============================================================

% #### Appendix ###############################################################################################################

\renewcommand{\theequation}{B.\arabic{equation}}
\setcounter{equation}{0}
\renewcommand{\thefigure}{B.\arabic{figure}}
\setcounter{figure}{0}
\renewcommand{\thetable}{B.\arabic{table}}
\setcounter{table}{0}

\section{Details of Graph Nodes Features}\label{appendix_feature_vector}

% ============================================================
\begin{table}[t]
  \centering
  \caption{Node feature vector ($D = 15$). ``$-1$'' indicates the feature does not apply; such dimensions are excluded from loss computation.}
  \renewcommand{\arraystretch}{1.4}
  \begin{tabular}{l l l l}
    \toprule\toprule
    \textbf{Index} & \textbf{Semantics} & \textbf{Entity Node} & \textbf{Fact Node} \\
    \midrule
    $x_v^{(0)}$               & Node-type indicator              & $0$                  & $1$ \\
    $x_v^{(1)}$               & Unique node index                & $\mathbb{Z}{\geq}0$  & $\mathbb{Z}{\geq}0$ \\
    \hdashline\\[-10pt]
    $x_v^{(2)}$               & Predicate type                   & $-1$                 & integer \\%$\{0,1,2,4,5,7,10\}$ \\
    $x_v^{(3)}$               & \textit{holds} value             & $-1$                 & $\{0,1\}$ \\
    $x_v^{(4)}$--$x_v^{(9)}$  & BGP configurations              & $-1$                  & integer / $-1$ \\
    $x_v^{(10)}$              & OSPF configurations             & $-1$                  & integer / $-1$ \\
    \hdashline\\[-10pt]
    $x_v^{(11)}$              & External AS entity              & $\{-1,1\}$            & $-1$ \\
    $x_v^{(12)}$              & Destination network entity      & $\{-1,1\}$            & $-1$ \\
    $x_v^{(13)}$              & Route reflector entity          & $\{-1,1\}$            & $-1$ \\
    $x_v^{(14)}$              & Router entity                   & $\{-1,1\}$            & $-1$ \\
    \bottomrule\bottomrule
  \end{tabular}
  \label{table_feature_vector}
\end{table}
% ============================================================

This appendix provides the complete specification of the node feature vector and fact node types introduced in Section~\ref{section_graph_representation}, including the full $D=15$ feature structure shared across entity and fact nodes (see Table~\ref{table_feature_vector}) and the argument structure of all fact node predicates (see Table~\ref{tab_fact_node_args}).

To support the batching of heterogeneous graphs during training in Section~\ref{section_proposed_algorithm}, both entity node and fact node share the same 15-dimensional feature structure, with dimensions that do not apply to a given node type set to $-1$. As summarized in Table~\ref{table_feature_vector}, the feature dimensions are organized into three groups. 
First, $x_v^{(0)}$ and $x_v^{(1)}$ are shared by both node types. Feature $x_v^{(0)} \in \{0, 1\}$ distinguishes entity nodes from fact nodes, and $x_v^{(1)}$ is a unique node index. 
Second, $x_v^{(2)}$ through $x_v^{(10)}$ are reserved for fact nodes. Feature $x_v^{(2)}$ encodes the predicate type, identifying which protocol relationship the fact node represents; feature $x_v^{(3)}$ is a \textit{holds} value, indicating whether that relationship is currently asserted in the network; and $x_v^{(4)}$ through $x_v^{(10)}$ carry protocol-specific configuration parameters, comprising six BGP route attributes and one OSPF link weight. 
Third, $x_v^{(11)}$ through $x_v^{(14)}$ are reserved for entity nodes, forming a binary indicator of the physical entity type: the dimension corresponding to the entity's type is set to $1$, and the remaining three are $-1$, identifying whether the entity is a router, RR, external AS, or destination network.

For each entity node $v \in \mathcal{V}_\mathrm{e}$, $x_v^{(0)} = 0$ and $x_v^{(1)}$ is the unique node index. The physical entity type is encoded by $x_v^{(11)}, \cdots, x_v^{(14)}$: $x_v^{(11)} = 1$ denotes an external AS, $x_v^{(12)} = 1$ a destination network, $x_v^{(13)} = 1$ a route reflector, and $x_v^{(14)} = 1$ a router. All other dimensions $x_v^{(2)}, \cdots, x_v^{(10)}$ are set to $-1$.

For each fact node $v \in \mathcal{V}_\mathrm{f}$, $x_v^{(0)} = 1$ is the node-type indicator and $x_v^{(1)}$ is the unique node index. $x_v^{(2)} \in \{0,1,2,4,5,7,10\}$ encodes the predicate type, corresponding to $\texttt{BGP\_route}(\cdot)$, $\texttt{connected}(\cdot)$, $\texttt{eBGP}(\cdot)$, $\texttt{fwd}(\cdot)$, $\texttt{iBGP}(\cdot)$, $\texttt{reachable}(\cdot)$, and $\texttt{trafficIsolation}(\cdot)$, respectively. Feature $x_v^{(3)} \in \{0,1\}$ is the \textit{holds} value, indicating whether the predicate is currently asserted.

% #### Appendix ###############################################################################################################

\footnotesize
\bibliographystyle{IEEEtran}
\bibliography{IEEEabrv, refs.bib}

% \vspace{-20 pt}

\begin{IEEEbiography}
[{\includegraphics[width=1in,height=1.25in,clip,keepaspectratio]{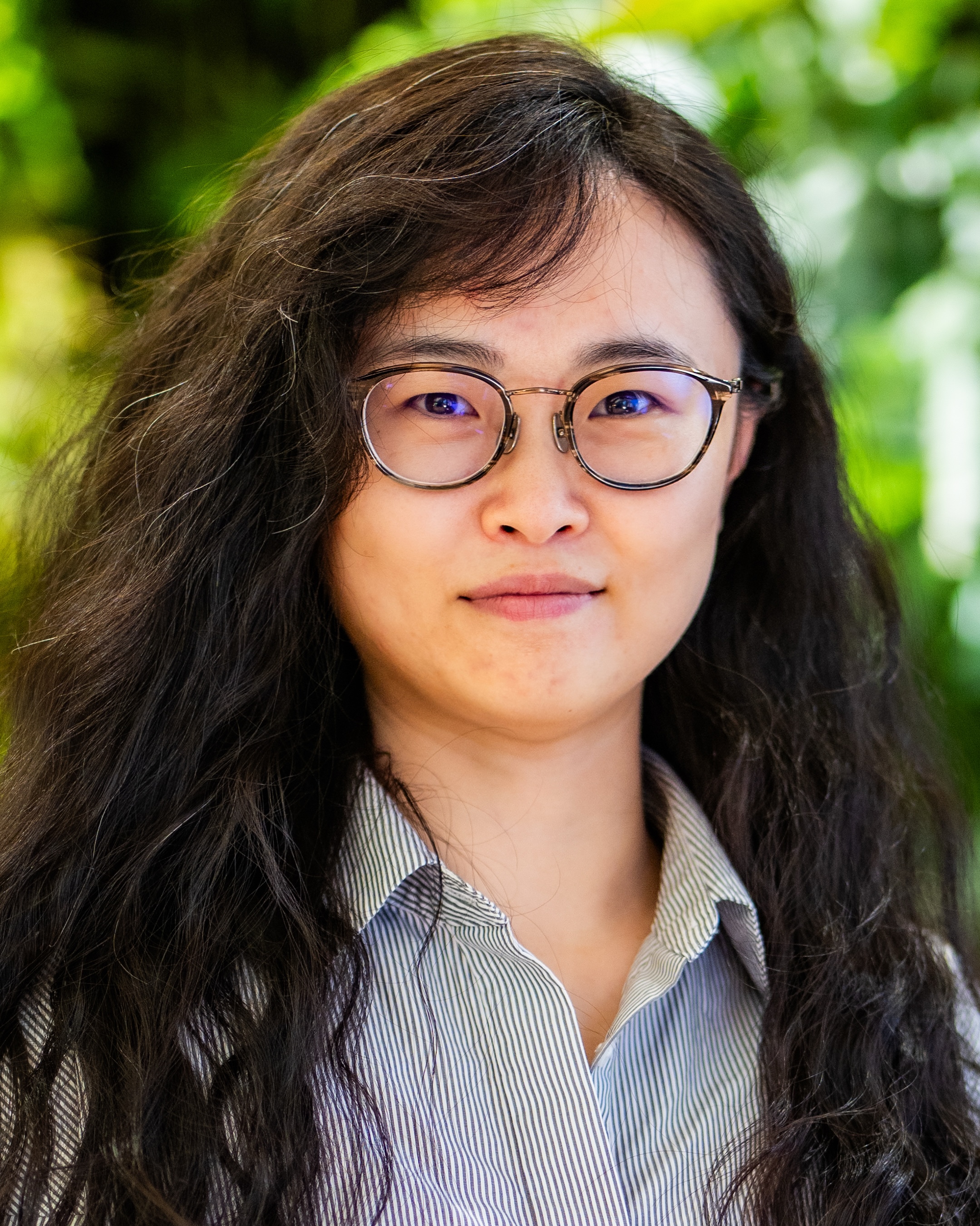}}] 
{Xin Hao} (Member, IEEE) received her Ph.D. degree from The University of Sydney in 2024. She is currently a Research Fellow with the Faculty of Engineering and IT at the University of Technology Sydney. 
% From 2024-2025, she was with the School of Information Technology at Deakin University. 
Her research interests include intelligent network resilience, graph neural networks, and multi-agent systems. She is a recipient of the USYD 2023 Faculty of Engineering PhD Completion Award, the 2023 Faculty of Engineering Research Scholarship, and the 2020 Faculty of Engineering Research Scholarship. She served as a session chair in the 2023 IEEE International Conference on Communications (ICC) workshop.
\end{IEEEbiography}

\vspace{-10 pt}

\begin{IEEEbiography}
[{\includegraphics[width=1in,height=1.25in,clip,keepaspectratio]{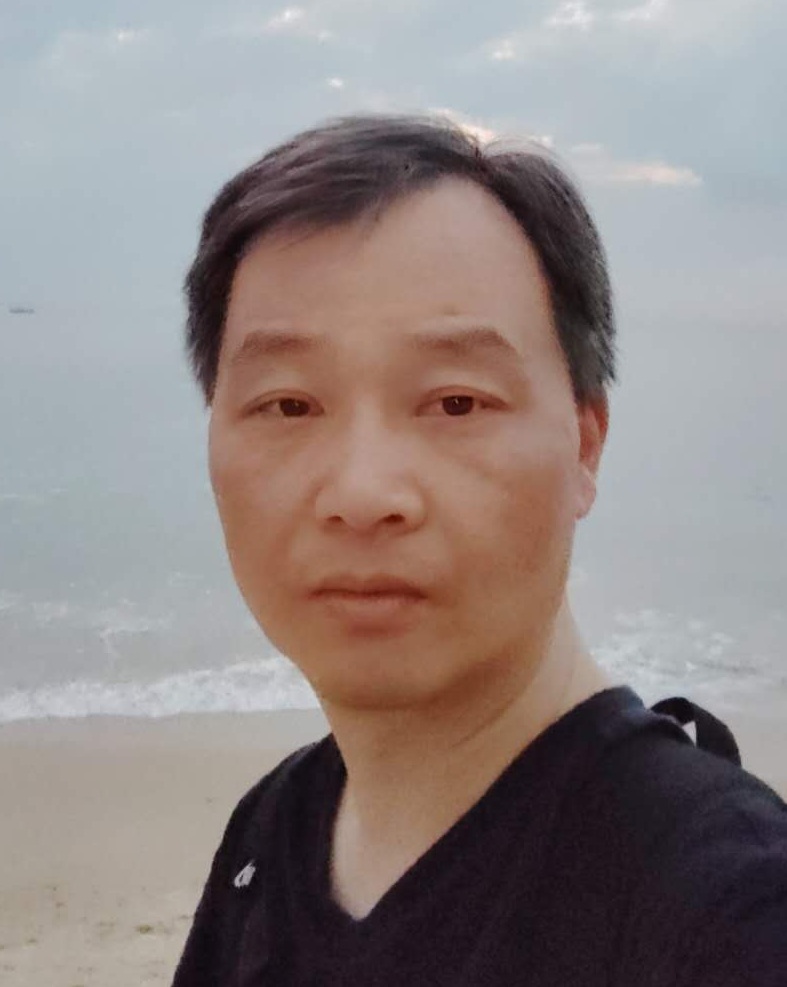}}] 
{Wei Ni} (Fellow, IEEE) received the B.E. and Ph.D. degrees in electronic engineering from Fudan University, Shanghai, China, in 2000 and 2005, respectively. He was a Deputy Project Manager with Bell Labs, Alcatel/Alcatel-Lucent from 2005 to 2008, Senior Research Engineer at Nokia from 2008 to 2009, and a Senior Principal Research Scientist and Group Leader with the Commonwealth Scientific and Industrial Research Organisation, from 2009 to 2025. He is currently the Associate Dean (Research) with the School of Engineering, Edith Cowan University, Perth, WA, Australia and an Adjunct Professor at the Telecommunication Research Institute, the University of Technology Sydney, NSW, Australia. He is also a Technical Committee Member with Standards Australia. His research interest include distributed and trusted learning with constrained resources, quantum Internet, and their applications to system efficiency, integrity, and resilience. He was the co-recipient of the ACM Conference on Computer and Communications Security 2025 Distinguished Paper Award, and four Best Paper Awards. He has been the Editor of IEEE TRANSACTIONS ON WIRELESS COMMUNICATIONS since 2018, IEEE TRANSACTIONS ON VEHICULAR TECHNOLOGY since 2022, IEEE TRANSACTIONS ON INFORMATION FORENSICS AND SECURITY and IEEE COMMUNICATION SURVEYS AND TUTORIALS since 2024, and IEEE TRANSACTIONS ON NETWORK SCIENCE AND ENGINEERING and IEEE TRANSACTIONS ON CLOUD COMPUTING since 2025. He was Chair of the IEEE VTS NSW Chapter from 2020 to 2022, Track Chair for VTC-Spring 2017, Track Co-chair for IEEE VTC-Spring 2016, Publication Chair for BodyNet 2015, and Student Travel Grant Chair for WPMC 2014.
\end{IEEEbiography}

\begin{IEEEbiography}
[{\includegraphics[width=1in,height=1.25in,clip,keepaspectratio]{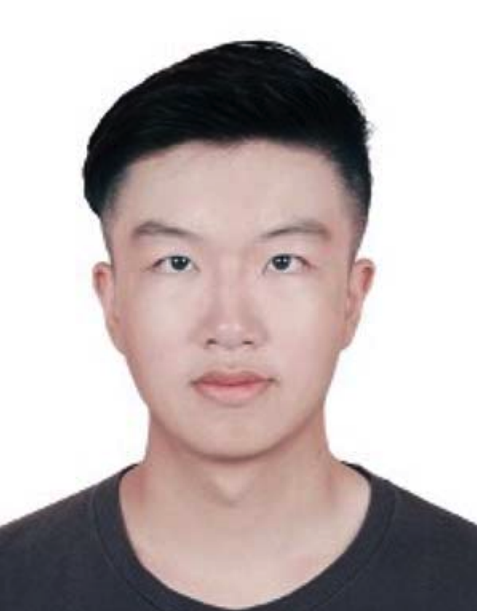}}] 
{Chenhan Zhang} (Member, IEEE) received the BEng (honours) degree from the University of Wollongong, Australia, in 2017, the MS degree from the City University of Hong Kong, Hong Kong, in 2019, and the PhD degree from the University of Technology Sydney, Australia, in 2024, where he was advised by Prof. Shui Yu. He is currently a postdoctoral research fellow with the University of Technology Sydney. His research interests include security and privacy in graph neural networks and trustworthy spatiotemporal cyber physical systems. He has been actively involved in the research community by serving as a reviewer for prestige venues, such as ICLR, IJCAI, INFOCOM, the IEEE Transactions on Dependable and Secure Computing, IEEE Internet of Things Journal, ACM Computing Survey, and IEEE Communications Surveys and Tutorials.
\end{IEEEbiography}

\begin{IEEEbiography}
[{\includegraphics[width=1in,height=1.25in,clip,keepaspectratio]{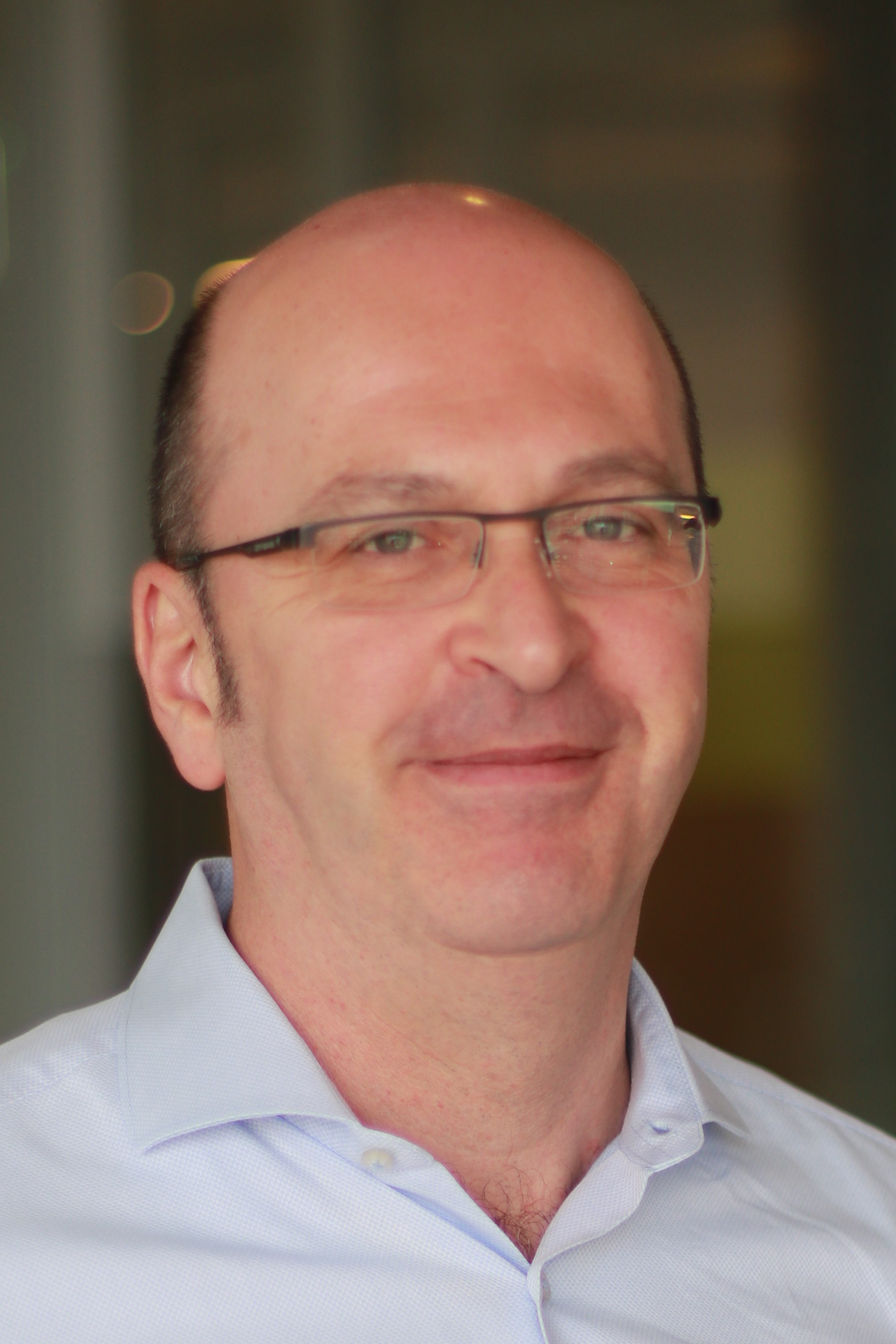}}] 
{Massimo Piccardi} (Senior Member, IEEE) received the M.Eng. and Ph.D. degrees from the University of Bologna, Bologna, Italy, in 1991 and 1995, respectively. He is currently a Full Professor of computer systems with the University of Technology Sydney, Ultimo, Australia. His research interests include natural language processing and machine learning and he has coauthored more than two-hundred papers in these areas. Prof. Piccardi is a member of the IEEE Computer and Systems, Man, and Cybernetics Societies, a member of the International Association for Pattern Recognition, and a member of the Association for Computational Linguistics.
\end{IEEEbiography}

\begin{IEEEbiography}
[{\includegraphics[width=1in,height=1.25in,clip,keepaspectratio]{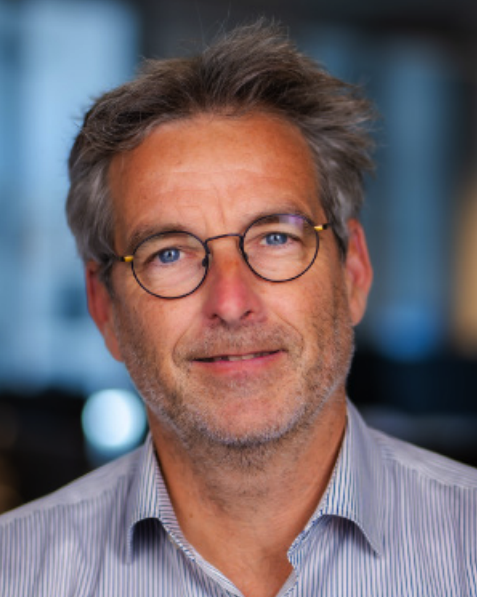}}] 
{Raymond Owen} (Member, IEEE) received the Bachelor of Engineering degree in electrical and electronic engineering from King’s College London, in 1992, and the Ph.D. degree in electrical and electronic engineering from the University of Birmingham, U.K., in 1995, where his Ph.D. research focused on modeling of high-frequency acoustic scattering from a rough surface. He is currently an Industry Professor with the Faculty of Engineering and IT, University of Technology Sydney (UTS). His research and commercial expertise span telecommunications resilience, digital modelling, fixed-line telecoms, network protocols, and wireless networking, particularly focusing on propagation. In addition to his academic role, he is a Partner with Tech Audit Partners, a U.S.-based consultancy specializing in resilience, and provides consultancy on network outages, their causes, and recovery strategies.
\end{IEEEbiography}

\end{document}